\documentclass[onecolumn,showpacs,prl,superscriptaddress]{revtex4}
\usepackage{graphicx}
\usepackage{color}

\usepackage{etoolbox}
\apptocmd{\thebibliography}{\raggedright}{}{}

\begin{document}

\title{Analysis of the evolution of the Sars-Cov-2 in Italy, the role of the asymptomatics and the success of Logistic model}

\author{Gabriele Martelloni}
\email{gabriele.martelloni@gmail.com}
\affiliation{Independent researcher}
\author{Gianluca Martelloni}
\email{gianluca.martelloni@unifi.it}
\affiliation{chemistry department- INSTM - via della Lastruccia 3-13, I - 50019 Sesto Fiorentino, Firenze, Italy.}

\begin{abstract}
In this letter we study the temporal evolution of the Sars-Cov-2 in Italy. The time window of the real data is between February 24 and March 25. After we upgrade the data until April 1.We perform the analysis with 4 different model and we think that the best candidate to describe correctly the italian situation is a generalized Logistic equation. We use two coupled differential equations that describe the evolution of the severe infected and the deaths. We have done this choice, because in Italy the pharyngeal swabs are made only to severe infected and so we have no information about asymptomatic people.  An important observation is that the virus spreads between Regions with some delay; so we suggest that a different analysis region by region would be more sensible than that on the whole Italy. In particular the region Lombardia has a behaviour very fast with respect to the other ones.  We show the behaviour of the total deaths and the total severe infected for Italy and five regions: Lombardia, Emilia Romagna, Veneto, Piemonte, Toscana. 
Finally we do an analysis of the peak and an estimation of how many lifes have been saved with the LockDown. 
\end{abstract}


\maketitle
In early December, or reasonably during November, the Sars-Cov-2, or commonly coronavirus, appear in Wuhan, China.\\
Coronaviruses are a large family of viruses known to cause diseases ranging from the common cold to more serious diseases such as Middle Eastern Respiratory Syndrome (MERS) and Severe Acute Respiratory Syndrome (SARS). Coronaviruses were identified in the mid-1960s and are known to infect humans and certain animals (including birds and mammals). The primary target cells are the epithelial cells of the respiratory and gastrointestinal tract and seven Coronaviruses are identified to be able to infect humans. They can cause common colds but also serious lower respiratory tract infections.\\
The virus causing the current coronavirus epidemic has been called "severe acute respiratory syndrome coronavirus 2" (SARS-CoV-2). This was announced by the International Committee on Taxonomy of Viruses (ICTV) which deals with the designation and naming of viruses (i.e. species, genus, family, etc.). The name is indicated by a group of experts specifically appointed to study the new coronavirus strain. According to this pool of scientists, the new coronavirus is the brother of what caused Sars (SARS-CoVs), hence the chosen name of SARS-CoV-2.\\
The disease caused by the new Coronavirus has a name: "COVID-19" (where "CO" stands for corona, "VI" for virus, "D" for disease and "19" indicates the year in which it occurred). The Oms Director-General Tedros Adhanom Ghebreyesus announced it on February 11, 2020, during the extraordinary forum dedicated to the virus.\\
The appearance of new pathogenic viruses for humans, previously circulating only in the animal world, is a widely known phenomenon (called spill over) and it is thought that it may also be at the basis of the origin of the new coronavirus (SARS- CoV-2). The scientific community is currently trying to identify the source of the infection.\\
On December 31, 2019, the Municipal Health Commission of Wuhan (China) reported to the Oms a cluster of cases of pneumonia of unknown etiology in the city of Wuhan, in the Chinese province of Hubei.\\
On January 9, 2020, the Chinese Center for Disease Prevention and Control (CDC) reported that a new coronavirus (initially called 2019- nCoV and now called SARS-CoV-2) has been identified as the causative agent and has been rendered publishes the genomic sequence.\\
On January 30, 2020, the World Health Organization declared that this epidemic represents an international public health emergency.\\
Oms on March 11, 2020 declared that COVID-19 can be defined as a pandemic.\\
After notification of the epidemic by China, Italy immediately recommended postponing unnecessary flights to Wuhan and, subsequently, with the spread of the epidemic, to all of China.
Consequently, the latter has canceled all flights from Wuhan.\\
On 30 January, the Minister of Health ordered the suspension of air traffic with the People's Republic of China, including the Special Administrative Regions of Hong Kong and Macao, on 30 January. The measure also applies to Taiwan.\\
In consideration of the Oms declaration of "International public health emergency", on January 31, 2020 the Council of Ministers declared a state of emergency as a consequence of the health risk associated with Coronavirus infection.\\
This disease does not save Italy that has become a protected area with the DPCM signed on the evening of 9 March by the Prime Minister, Giuseppe Conte, who has extended the restrictive measures already applied for Lombardy and the 14 northern provinces most affected by the coronavirus infection to the whole national territory. The new action comes into force on March 10 and will take effect until April 3. Among the main innovations: it limits the movement of people, blocks sporting events, suspends teaching activities in schools and universities throughout the country until April 3.\\
With the new ordinance of 22 March 2020 issued by the Minister of Health and the Minister of the Interior, from 22 March people are prohibited from moving or moving with public or private trasportation in a municipality other than that in which they are located, except for proven work needs, of absolute urgency or for health reasons.\cite{salute.gov.it}\\
In this scenario let's try to analyze the time evolution of the Sars-Cov-2 in Italy with growth patterns. Many growth models have been very recently applied to study the Covid-19 infection\cite{LBAGZ, CLG, Casto1, Fenga, FP, AG, Betal, Casto2}. Two very famous examples of these models are described by the Logistic\cite{Logistic} and the Gompertz\cite{Gompertz} equations. The Logistic law has been applied in population dynamics, in economics, in material science and in many other sectors, while the Gompertz one describes tumor growth, kinetics of enzymatic reactions, oxygenation of hemoglobin, intensity of photosynthesis as a function of CO2 concentration, drug dose-response curve, dynamics of growth, (e.g., bacteria, normal eukaryotic organisms). \\
The Logistic behaviour assumes that growth stops when maximum sustainable population density is reached through the carrying capacity K that depends on the environmental conditions. For example the ordinances of the Prime Minister G.Conte, the people's hygiene habits are encoded in the carrying capacity K.\\
Whereas the Gompertz equation takes into account the aging of the population through a lower reproductive capacity over time; not knowing yet the characteristics of the virus we thought it could be a correct choice.\\
Now we describes the regional and national choices to contain the epidemic and and consequently our strategy.\\
At first in Italy, pharyngeal swabs were initially made only at seriously ill people. This choice, extremely correct in safeguarding the hospital structure (at least apparently), is wrong for a study of the disease data. We explain better: if we believe that the Case Fatality Ratio (CFR) is $1\%$ \cite{Imperial} the number of deaths is the only data extremely relevant. From the extrapolation of total number of infected \footnote{Total number of infected is simply total number of deaths x 100 or if we believe that in Italy the CFR is $1,2\%$ (for the high number of ancient people) simply total number of deaths x 83,3.} we observe that the number of infected detected is only $10\%$ of the total. This means that in Italy about $90\%$ of the sick are not detected. \\
Having no information, therefore, on asymptomatic and infected with mild symptoms, we decided to study the population of the severe infected with 2 equations: one for the severe infected and another coupled for the deaths, obviously the latter are related to the former. So we implement a Logistic model that take into account a possibility to consider the contribute of asymptomatic people in the evolution of the Sars-Cov-2 by means of additional constant and an equation of proportionality between infected and deaths. We consider also an interpolating between Logistic and Gompertz Model. \\
We show that the first scenario is more optimistic and faster respect to the second one.\\
Finally we observe the data of each region separately, noting that is present a delay in the evolution of the virus between the various regions and that each region has its own trend. 
In particular, Lombardia has the fastest trend and Veneto the slowest one. We conclude with the observation that Veneto's strategy of performing swabs even to asymptomatic potential seems to have successful, as can be seen from the estimate of the constant term of the logistic model.

\section{The Logistic Model}
A possible scenario of the evolution of the evolution of Sars-Cov-2 virus in Italy is described by the logistic differential equation. The model is represented by the following equation
\begin{equation}
\frac{d I(t)}{d t}=r_{0} I(t) (1-\frac{I(t)}{K}), 
\end{equation}
where $I(t)$ is the number of infected and the time $t$ is the integration variable. The constant $r_{0}$ is the growth rate of this particular population (the Sars-Cov-2) and $K$ is the carrying capacity.\\
The growth rate $r_{0}$ is typical of the population because it is linked to the reproductive mechanism of the virus while the carrying capacity K depends on the environmental conditions. For example the government action of the Prime Minister G.Conte (Locked down(LD)), the people's hygiene habits are encoded in the carrying capacity K. \\
However we change this equation adding a constant term $A$ 
\begin{equation}
\frac{d I}{d t}=r_{0} I (1-\frac{I}{K})+A
\end{equation}
that represents the contribution of asymptomatic people. We explain our idea for the Italian situation. Studying \cite{Imperial} we assume that the CFR is $1\%$. Consider, for example, the situation of the March 19 in Italy
\begin{itemize}
\item 3405 deaths,
\item 41035 infected (severe cases, which is presented the necessity to verify the positivity to avoid the collapse of the healthcare facility),
\item 340500 estimated data of deaths $+$ severe, symptomatic and asymtomatic cases.
\end{itemize}
This means that we have a bath of 300000 infected that can produce severe cases. We consider for simplicity that this growth rate is constant.  \\
The exact solution for this equation is well-known
\begin{equation}
I(t)=\frac{K}{2}+\frac{\sqrt{K(4 A + K r_{0})}\tanh(\sqrt{K(4 A + K r_{0})}t)}{2\sqrt{r_{0}}}
\end{equation}
and the limit for $t->\infty$
\begin{equation}
I(\infty)=\frac{K}{2}(1+\frac{1}{r_{0}}\sqrt{\frac{4r_{0}A}{K}+r_{0}^{2}})
\end{equation}
represents the maximum possible number for this population.\\
After the March 23 we think it will be necessary recalibrate A with the data of the Civil Protection. Why? Because 14 days after the LD the incubation period is over and we suppose that this action destroys possible places for the propagation of the virus. Finally we have
\begin{eqnarray}
\frac{d I}{d t} &=& r_{0} I (1-\frac{I}{K})+A \qquad\mbox{t=March 23}, \\
\frac{d I}{d t} &=& r_{0} I (1-\frac{I}{K})+A_{LD} \qquad\mbox{t $>$ March 23}.
\end{eqnarray}
where $A_{LD}<A$ has the meaning that the LD restricts the possibility of contagion through asymptomatic in individual houses or just on the job places.\\
After 11 days the possibility to have symptoms are $2,5\%$ \cite{incubation} then also March 20 is a good candidate to impose with the data $A_{LD}<A$. We can also imagine the possibility that the LD would be so good that $A_{LD}<0$.\\
We remark that the exact date of the decrease in the infected is so difficult because the results of the test is not always referred to the same day but we can have a delay for the incredible number of pharyngeal swabs.\\
In conclusion we think that a day between March 20 and March 23 is a good candidate to the drop in infections.\footnote{Indeed we observe the March 22 as the date of the drop.}\\
Now we consider the second population of this model, the deaths. We couple the infected $I$ with the deaths $D$ in the following equation:
\begin{equation}
\frac{d D(t)}{d t}=K_{1} \frac{d I(t-t_{d})}{d t}. 
\end{equation}
After a simple integration we obtain
\begin{equation}
D(t)=K_{1}I(t-t_{d})+D(t_{1})-I(t_{1}-t_{d}). 
\end{equation}
where $t_{1}$ is the first integration extreme, reasonably the February 24.\\
So we fix to have the same growth for the severe cases and the deaths; $t_{d}$ represents the interval between the appearance of symptoms and the deaths. From \cite{DT} we estimate $t_{d}=13$ days, but we observe from numerical aspects that $t_{d}=4$. A brief comment: this reflects the italian choice to do swabs only to severe cases, and not at the very beginning of the disease.

\section{The Gompertz Model}
The second scenario of the evolution of Sars-Cov-2 virus in Italy is described by the Gompertz differential equation. The model is represented by 
the following equation
\begin{equation}
\frac{d I(t)}{d t}=-r_{g} I(t) \ln(\frac{I(t)}{K_{g}}), 
\end{equation}
where $I(t)$ is the number of infected and the time $t$ is the integration variable. The constant $r_{0}$ is the growth rate of the virus Sars-Cov-2 and $K$ is the carrying capacity.\\
This model takes into account the aging of the population through a lower reproductive capacity over time, in other words we consider the virus less contagious over time. \\
The solution of the previous equation give us the Gompertz law
\begin{equation}
I(t)=K_{g}e^{e^{-r_{g}t}}.
\end{equation}
As in the logistic study we compute the limit or $t->\infty$ and we obtain
\begin{equation}
I(\infty)=K_{g},
\end{equation}
this represents the asymptotic term of the population, defined by the resources available in the environment, in some sense also in this model we can encoded the human external action.\\
We choose to implement also in this model a constant term that we think that describes the role of the asymptomatic people. So we study numerically this generalized Gompertz equation
\begin{equation}
\frac{d I(t)}{d t}=-r_{g} I(t) \ln(\frac{I(t)}{K_{g}})+A.
\end{equation}
\section{The Interpolating Model}
From the study of the data, as we will see later in this letter, it's clear that the logistic scenario is very optimistic respect to the natural disaster described by the Gompertz model. Our idea is to create a model with a new parameter that enables to interpolate between the two regimes.\\
So we take into account the following differential equation
\begin{equation}
\dot{I}=q r_{0} I (1-\frac{I}{K_{l}}) - (q-1) r_{0} I\ln(\frac{I}{K_{g}})+A, 
\end{equation}
where $r_{0},K_{l}$ are respectively the growth rate and the carrying capacity of the Logistic model, and $r_{0},K_{g}$ for the Gompertz one. Obviously the time $t$ is always the integration variable. The parameter q has range between $0$ and $1$, and obviously for q=0 we obtain the Gompertz behaviour and for q=1 the Logistic one.
\section{The situation in Italy}
From the numerical simulations, performing 200 stochastic runs of the Gillespie direct method algorithm applied to the logistic equation, we obtain the following parameters for the Italy situation described by fig.1:
\begin{equation}
r_{0}=0.200\pm 0.001, K=110950\pm20\quad\mbox{and}\quad A=49\pm3, 
\end{equation}
and
\begin{equation}
K_{1}=0.14\quad\mbox{and}\quad t_{d}=4 .
\end{equation}
The errors are estimated at 5 $\sigma$, or about $5\%$.\\
\begin{figure}[hbt!]
\center\includegraphics[width=0.35\textwidth]{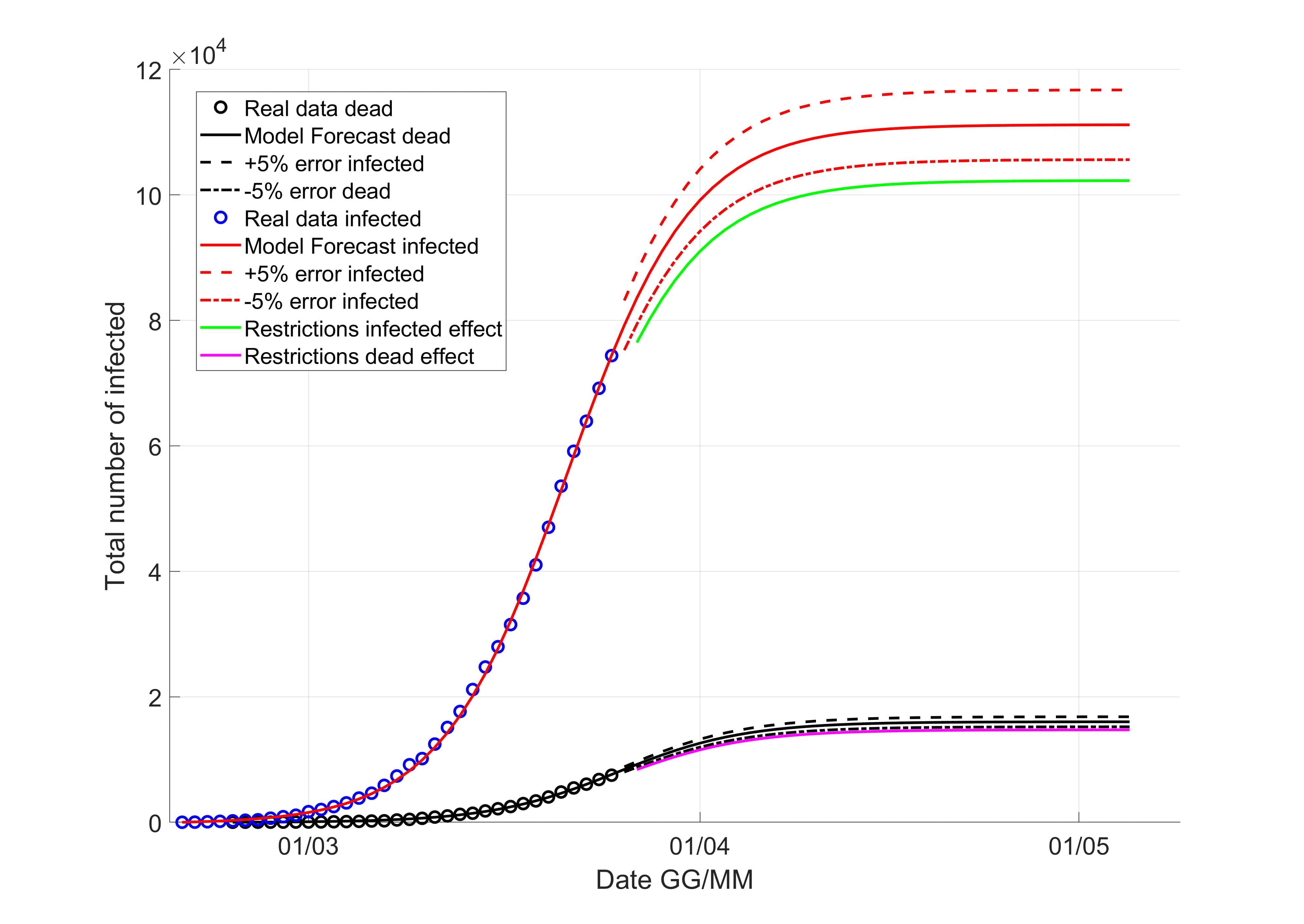}
\caption{The logistic curve of severe infected and deaths in Italy.}
\label{dataCP1}
\end{figure}
\begin{figure}[hbt!]
\center\includegraphics[width=0.35\textwidth]{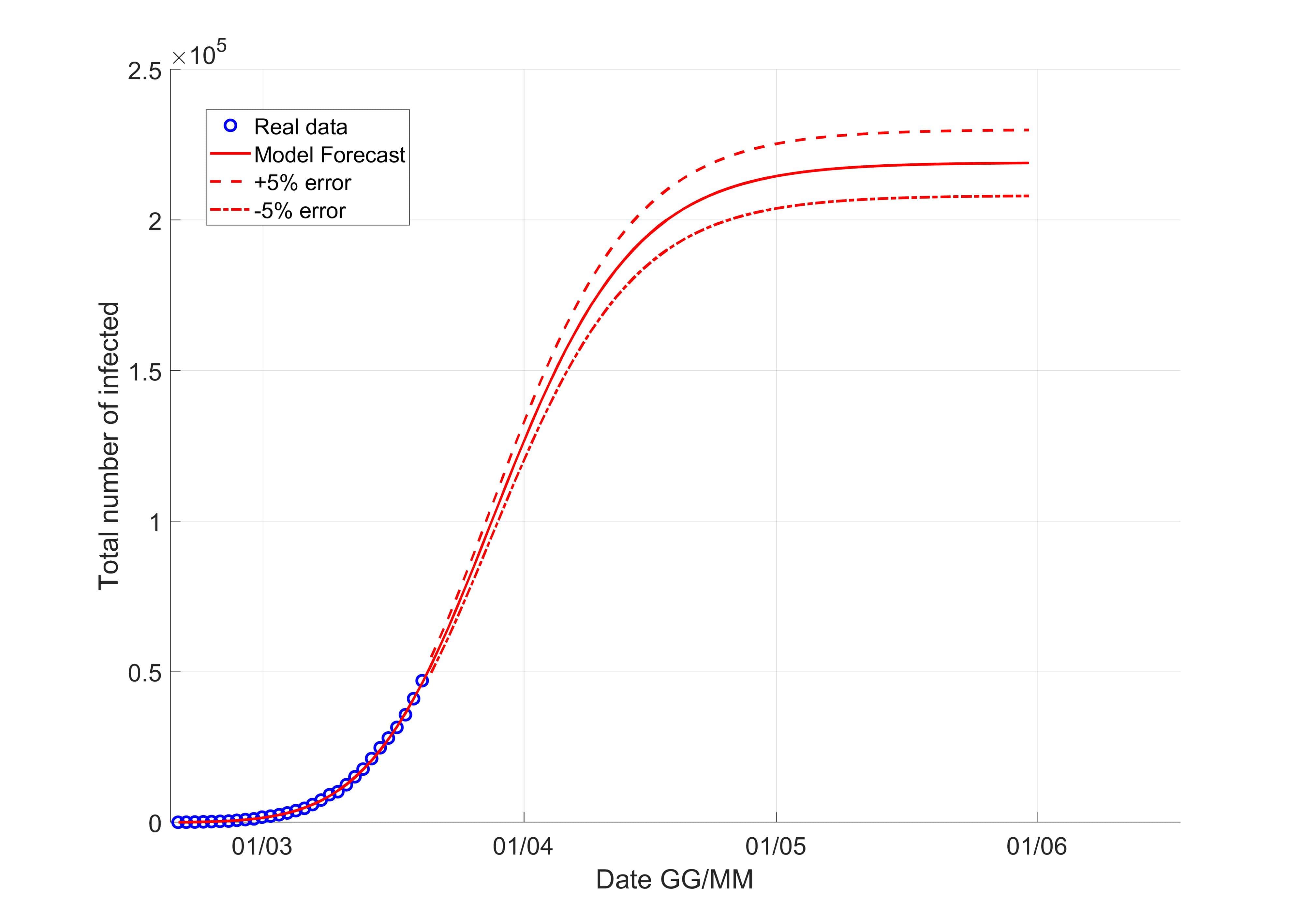}
\caption{The interpolating curve of severe infected in Italy at March 21 with $q=0.7$.}
\label{dataCP2}
\end{figure}
\begin{figure}[hbt!]
\center\includegraphics[width=0.35\textwidth]{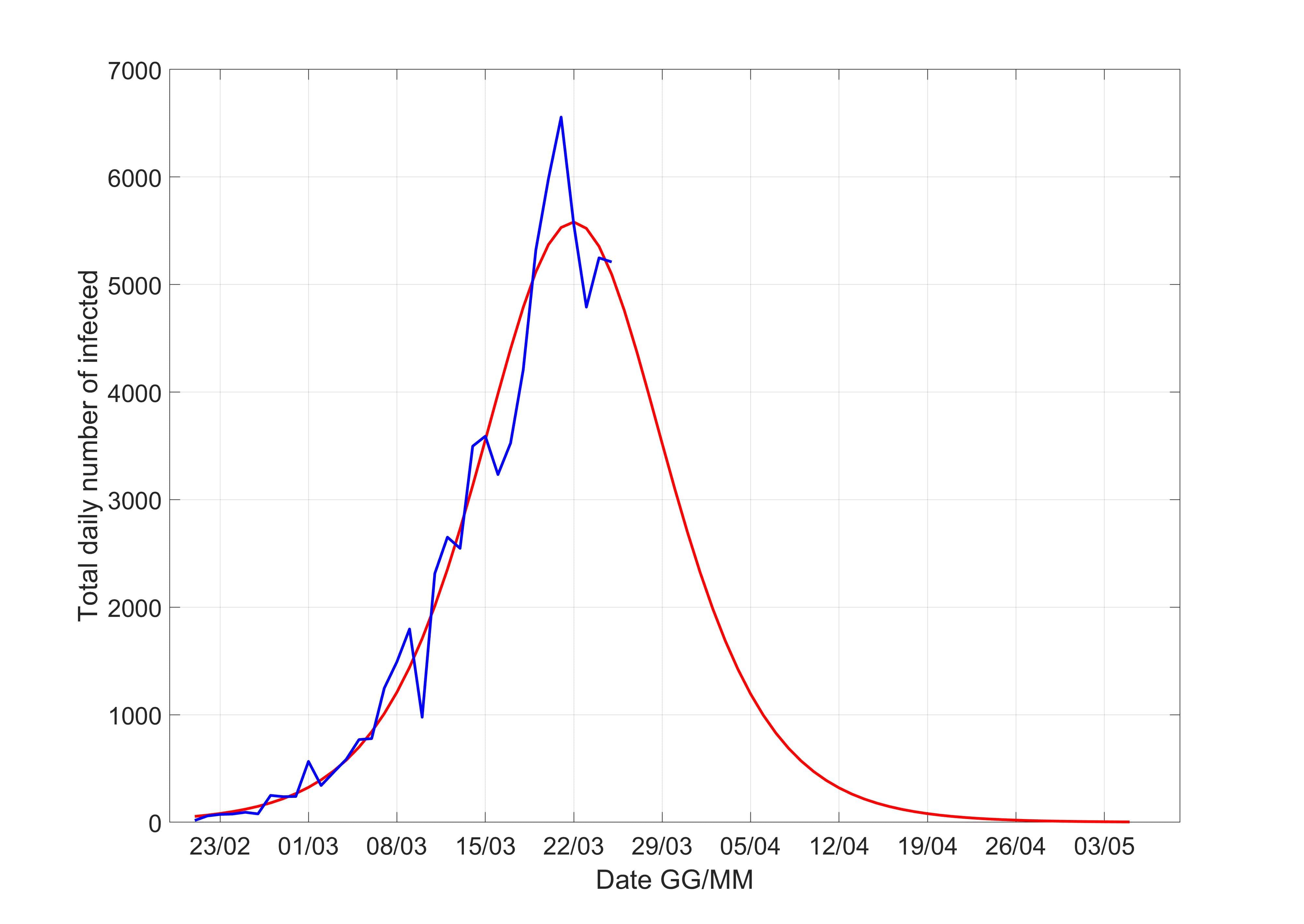}
\caption{The peak at 21 March and the end of the epidemic.}
\label{dataCP3}
\end{figure}

With the interpolating model at 21 March in Fig. 2 we observe how the situation could become particularly dramatic, especially as regards the duration of the epidemic (the curve becomes flat in late May). However, it becomes comforting with the data of March 25; the two graphs overlap and from the numerical simulations we find q = 1. This means that environmental factors, or LD, prevail over virus growth factors. 15 days after the LD we therefore started to see concrete effects, and this is compatible with the incubation times of the virus.\\ It therefore seems obvious to study the evolution of the virus with a Logistic model when a Government applies LD. \\
In conclusion, we expect around 111,000 diagnosed patients and around 16,000 deaths. This, considering a CFT of $1.2\%$, would lead to a total of about 1334,000 infected. \\
In Fig. 3 we see that the peak is near 21-22 March and the end of the epidemic at the early May.
\subsection{Updating of Italy situation at 27/03/2020}

In this section we update the Italy situation according to the calibration performed using the real data of Civil Protection at 27/03/2020. In Fig. 4 we report the previsions of infected and deaths: with the assimilation of two new data for the calibration of the model, the total number of infected goes to about 124800 and the deaths to about 17800. Observing Fig. 5, we can note how the curve of daily total infected changes from day to day according to the acquisition of a new data, but the right tails of the considered curves are similar. This suggests that the end of contagion is confirmed, according to our study, at the end of April or early May. Finally, in Fig. 6 we show the correlation, according to the delay, between the temporal series of total infected and deaths. The plot represents the real data of deaths and the real ones of the total infected shifted of 4 days and rescaled according to the minimum difference between them, i.e., we search the best correlation between the deaths data and the infected shifted forward according to Eq.7 (the deaths are delayed compared to infected). By means of this analysis we found the correct delay and we calibrate the parameter $K_{1}$ of Eq.7 that represents the rate between the derivatives of infected and deaths (the value of $K_{1}$, update at 27/03/2020, is 0.1429).

\begin{figure}[hbt!]
\center\includegraphics[width=0.35\textwidth]{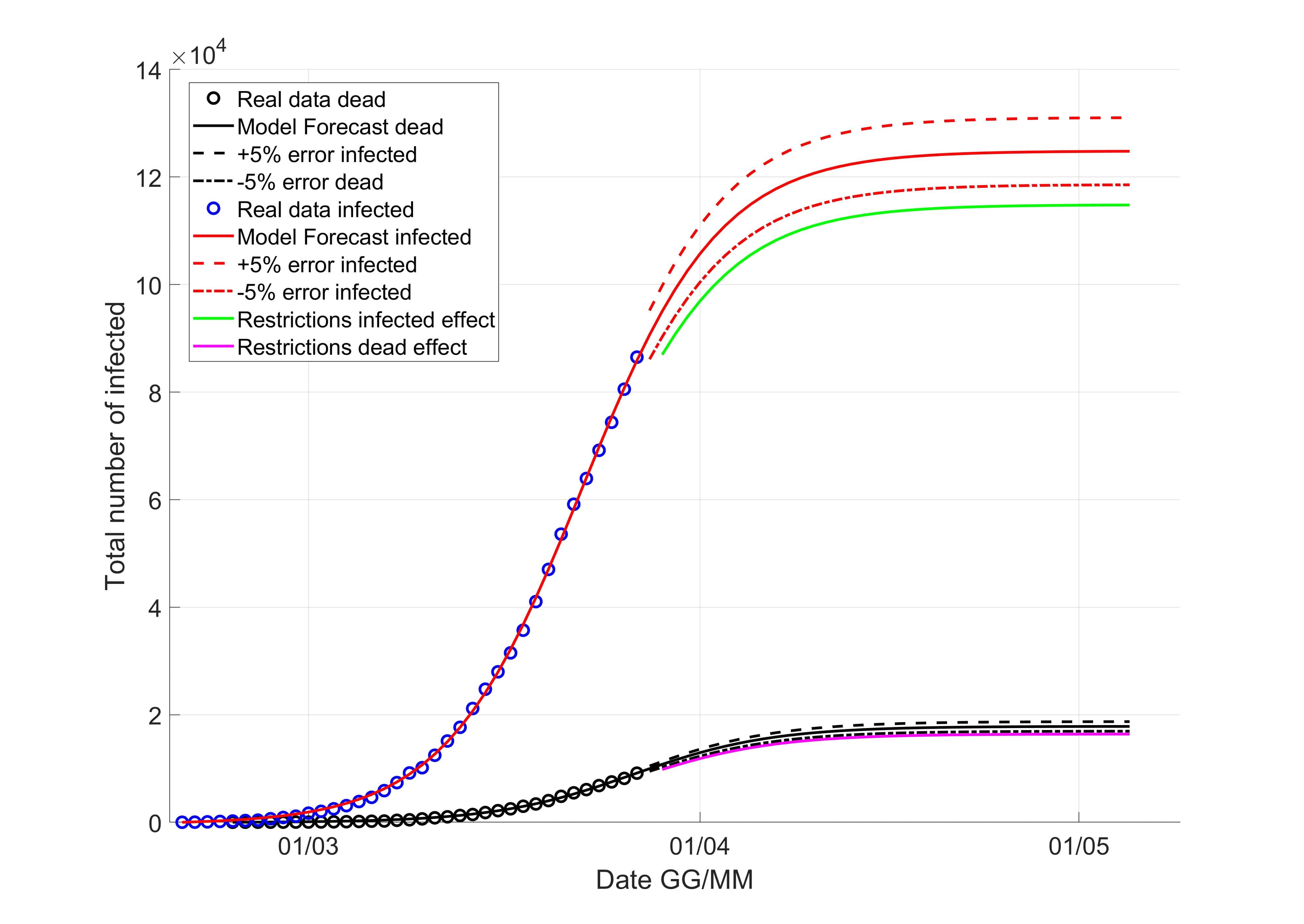}
\caption{The logistic curve of severe infected and deaths in Italy calibrated according to the data of 27/03/2020.}
\label{dataCP4}
\end{figure}
\begin{figure}[hbt!]
\center\includegraphics[width=0.35\textwidth]{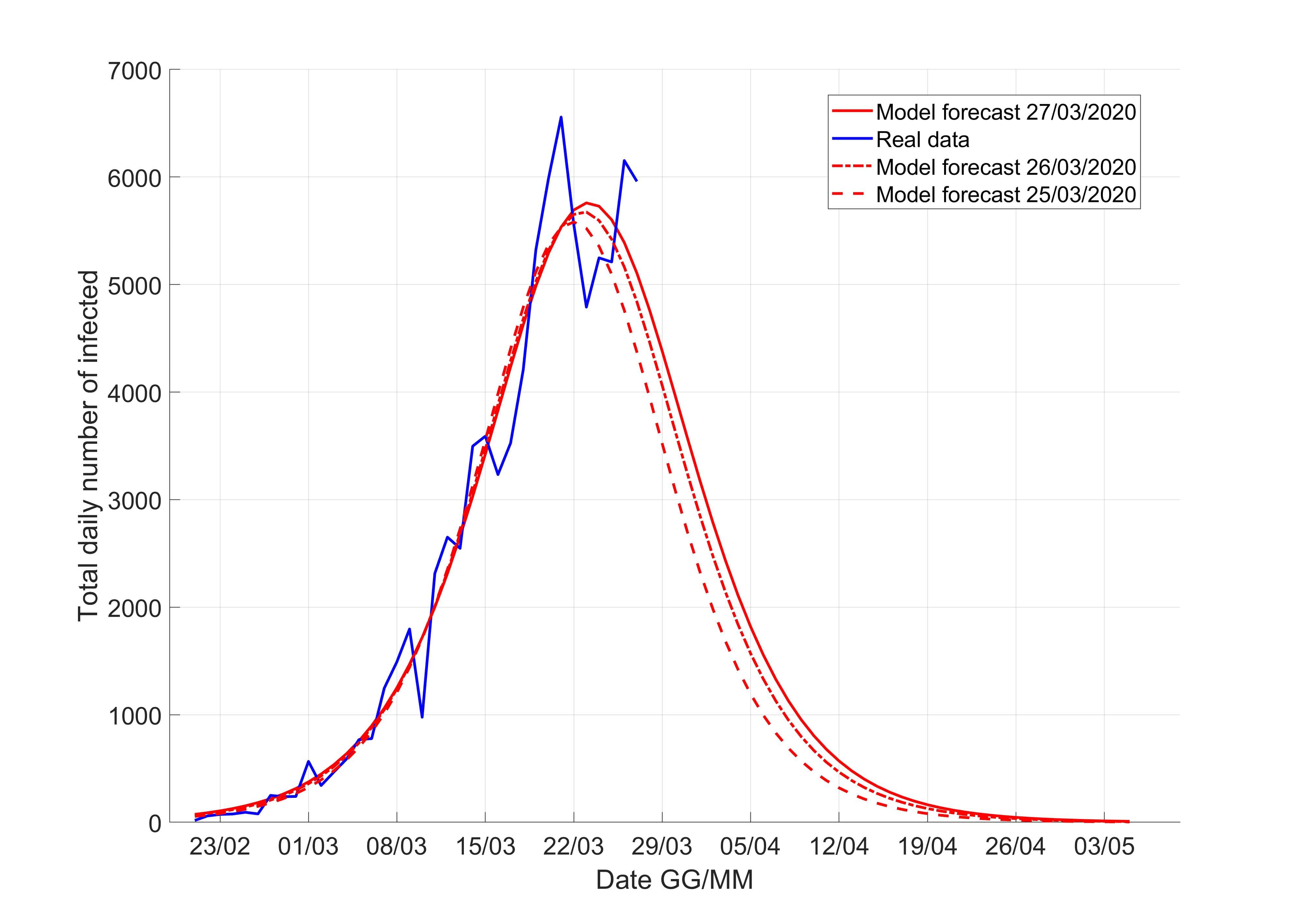}
\caption{Comparison between the daily total infected prevision in the last three days 25-26-27/03/2020.}
\label{dataCP5}
\end{figure}
\begin{figure}[hbt!]
\center\includegraphics[width=0.35\textwidth]{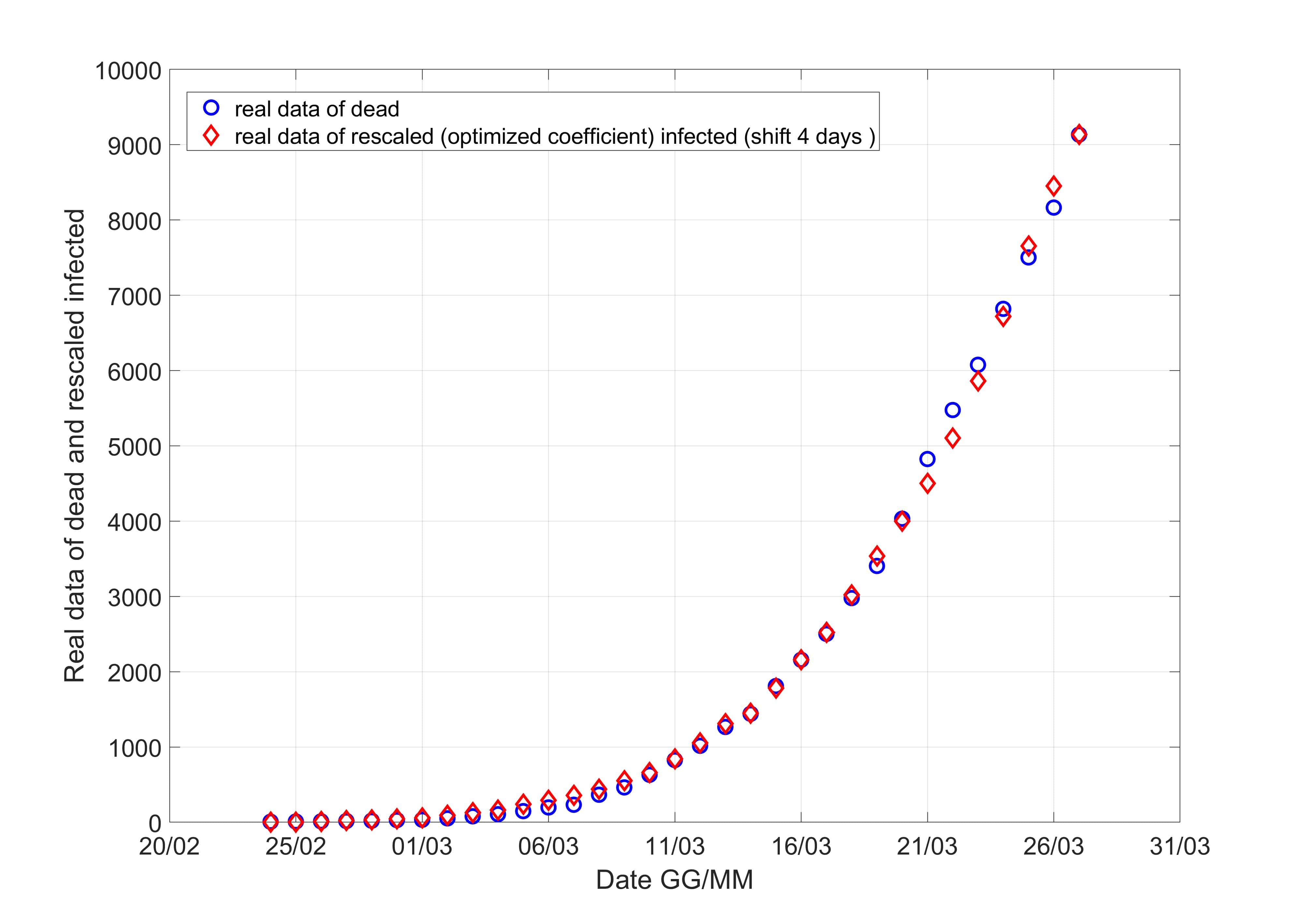}
\caption{Analysis of delay between infected and dead: for Italy delay is 4 days; the plot represent the real data of dead (blue circles) and the real data of total infected (red diamonds) shifted of 4 days and rescaled according to the minimum difference.}
\label{dataCP6}
\end{figure}

\section{Upgrade at April 1 and the generalized Logistic model: considerations about a peak marker}
We consider the following generalized Logistic equation
\begin{equation}
\frac{d I(t)}{d t}=r_{0} (I(t)^{\alpha}-\frac{I(t)^{2}}{K})+A, 
\end{equation}
where $\alpha$ is an experimentally parameter to fit better the initial power law climb. \\
We show the trend of the severe infected in Fig.7, Fig.8 and in Fig.9 respectively with the $\alpha$-model, the Logistic calibrated with data at 29 March and with the $\alpha$-model calibrated at 1 April.\\
\begin{figure}[hbt!]
\center\includegraphics[width=0.35\textwidth]{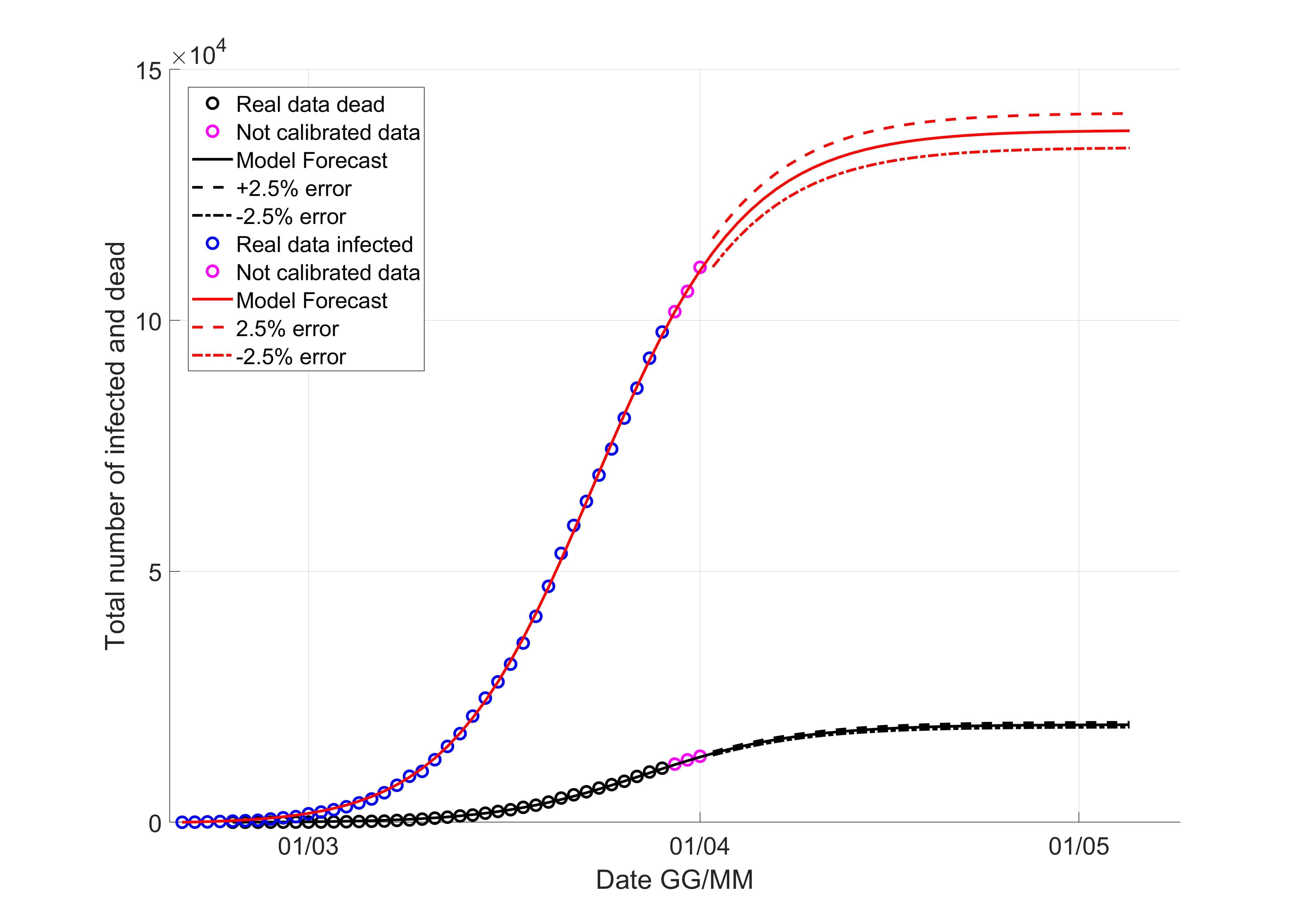}
\caption{The curve of deaths and severe infected for Italy with the $\alpha$-model calibrated at 29 March with data until the 1 April.}
\label{dataCP7}
\end{figure}
\begin{figure}[hbt!]
\center\includegraphics[width=0.35\textwidth]{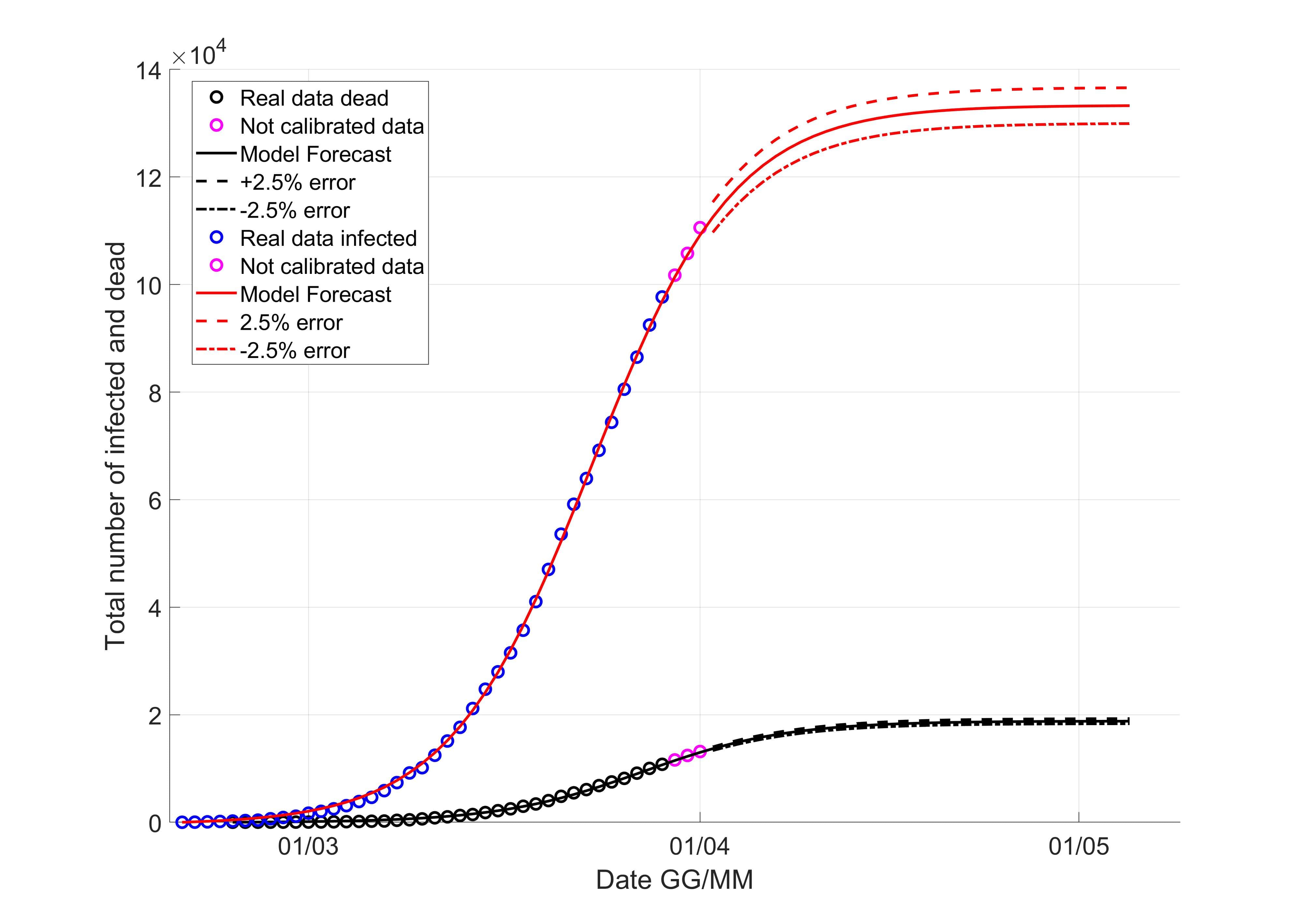}
\caption{The curve of deaths and severe infected for Italy with the Logistic model calibrated at 29 March with data until the 1 April.}
\label{dataCP8}
\end{figure}\begin{figure}[hbt!]
\center\includegraphics[width=0.35\textwidth]{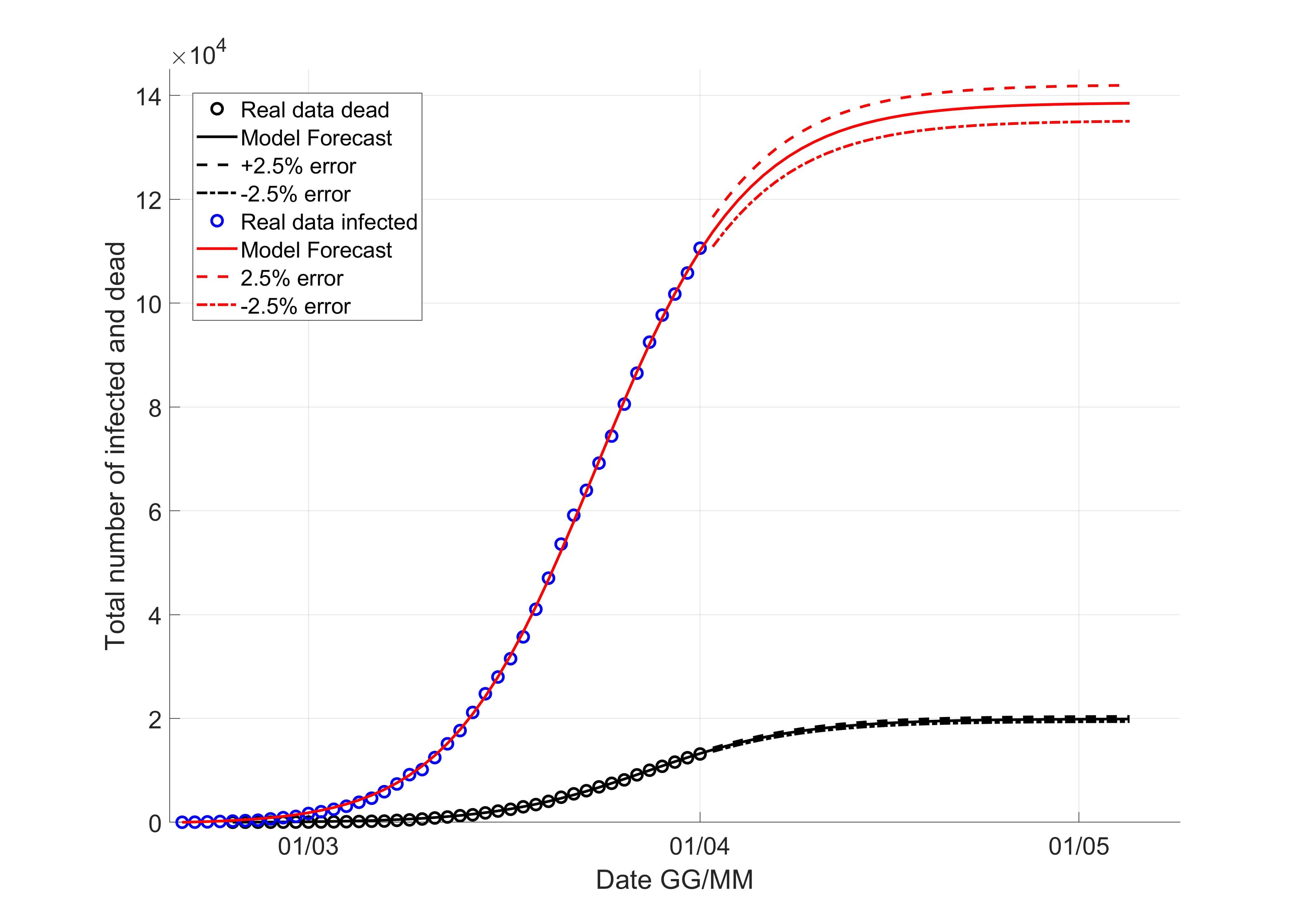}
\caption{The curve of deaths and severe infected for Italy with the $\alpha$-model calibrated at 1 April.}
\label{dataCP9}
\end{figure}
We observe as fitting better the initial power law the estimated number of severe infected and deaths increases of $2\%$, so we choose to use the $\alpha$-model to estimated the total number of severe infected and deaths. We obtain
\begin{equation}
\nonumber
I(\mbox{end})=138500,
D(\mbox{end})=19900,
\end{equation}
with the following values of the parameters
\begin{equation}
r_{0}=0.173\pm 0.004, K=138400\pm3500\quad\mbox{and}\quad A=79\pm2. 
\end{equation}
A brief comment, as we expect the growth coefficient $r_{0}$ is falling and the carrying capacity is obviously increased simply because the number of infected has grown. It is interesting that the contribution $A$ of asymptomatics is also increased. We can explain this by thinking that asymptomatics remain the most important sources of contagion in the single houses of italian people during the LD.
We observe as in a week after the peak the estimated number of severe infected and deaths are increased of $20\%$, but in the last 4 days the trend is in accordance with the data, both numerically and in forecasting the end of the epidemic, in particular the end date is stable from March 27. For this stability we decide to estimate the errors at 2,5 $\sigma$, i.e. $2,5\%$.\\
A possible marker of the presence of the peak could be the following quantity:
\begin{equation}
P_{\mbox{marker}}=\frac{I(t_{i})}{S(t_{i})},
\end{equation}
where $I(t_{i})$ is the total number of infected at the day $t_{i}$ and $S(t_{i})$ is the total number of swabs at the same day. As we see in Fig. 10 $P_{\mbox{marker}}$ has a peak at 24 March, which is basically in agreement with our prediction. The graph of Fig. 10 has been reproduced with a moving average. 
\begin{figure}[hbt!]
\center\includegraphics[width=0.35\textwidth]{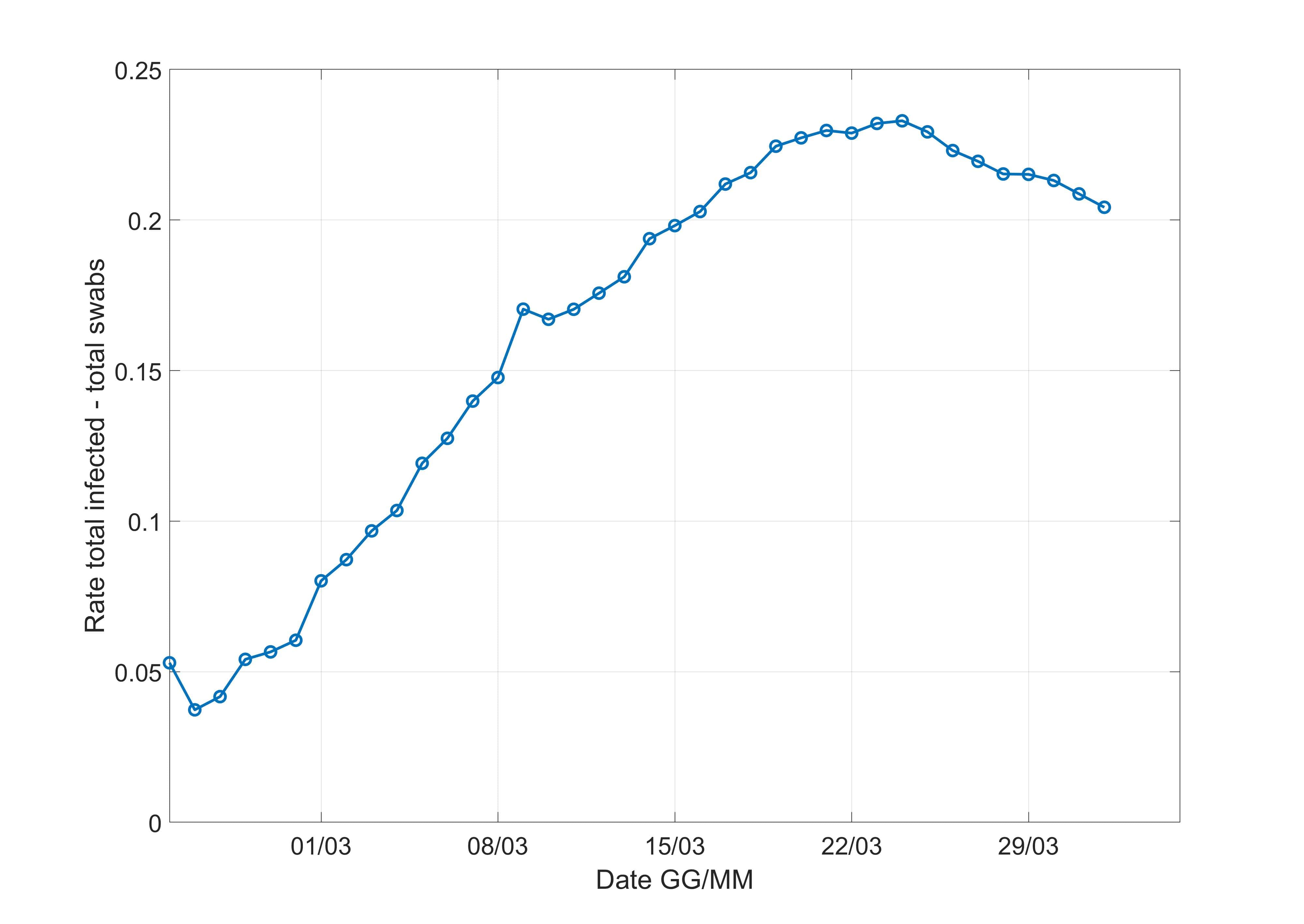}
\caption{The data of the Civil Protection for $P_{\mbox{marker}}$ until 1 April}
\label{dataCP10}
\end{figure}
\begin{figure}[hbt!]
\center\includegraphics[width=0.35\textwidth]{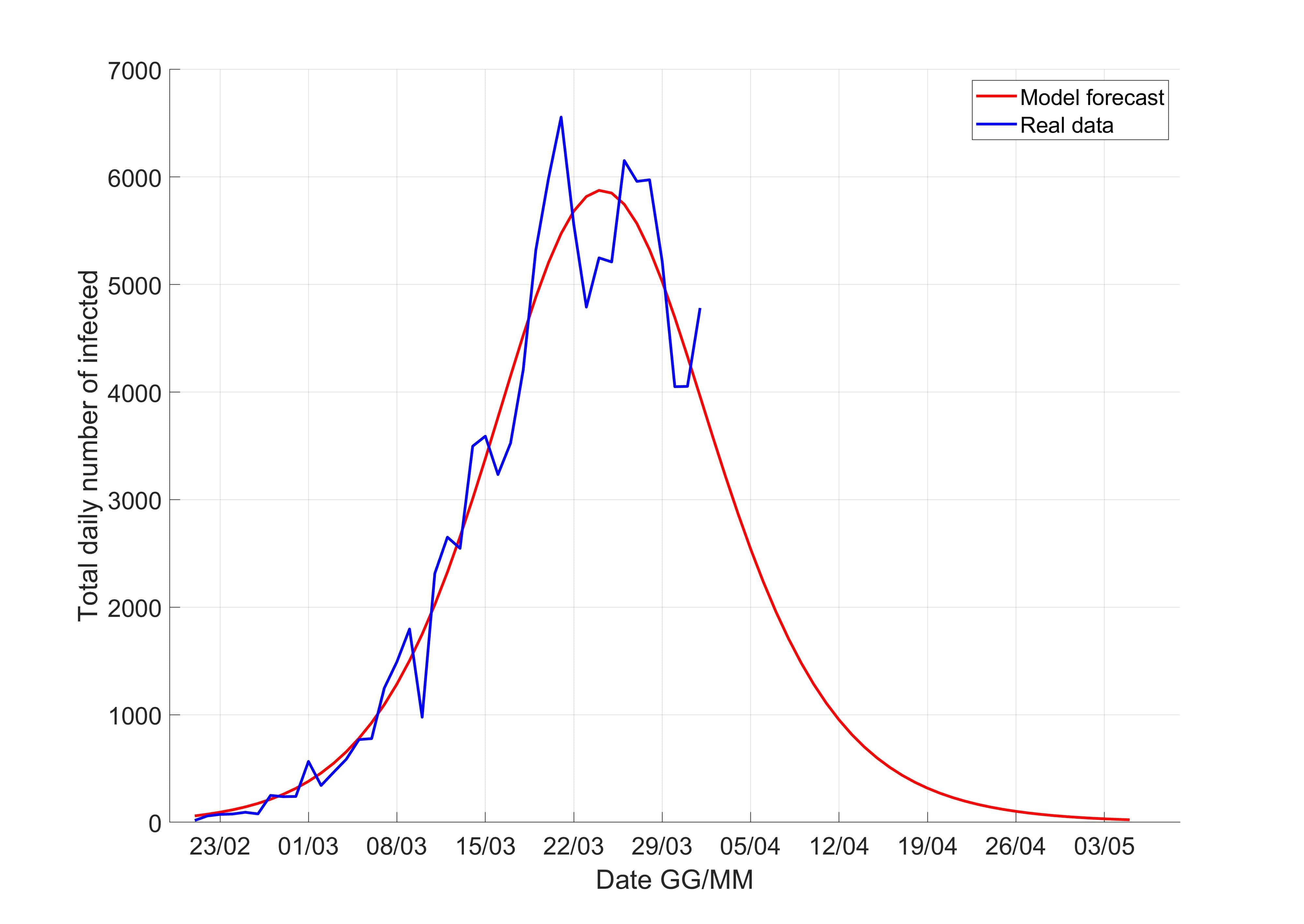}
\caption{The data of the Civil Protection and the theoretical curve of severe infected for single day.}
\label{dataCP11}
\end{figure}
\begin{figure}[hbt!]
\center\includegraphics[width=0.35\textwidth]{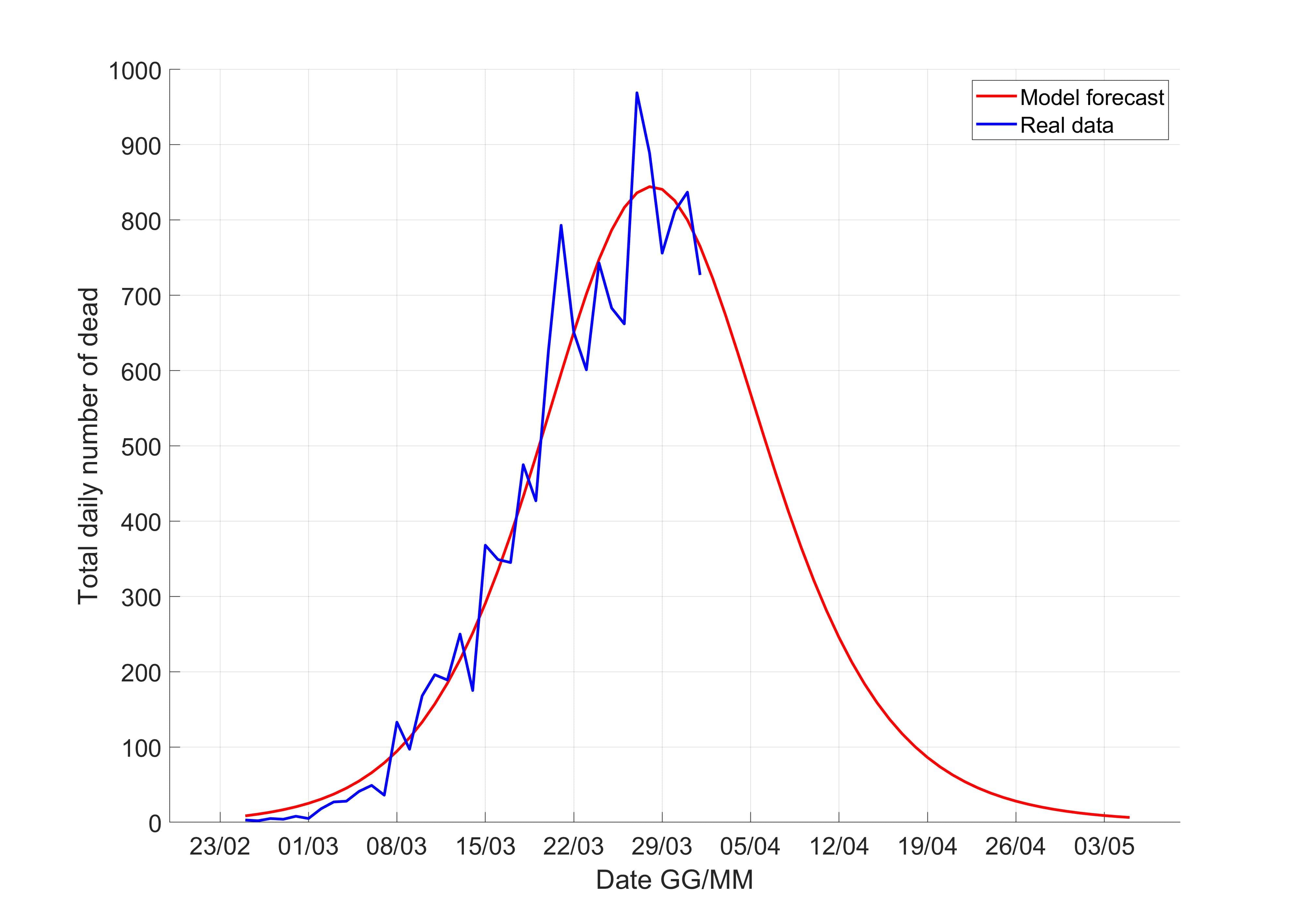}
\caption{The data of the Civil Protection and the theoretical curve of severe infected for single day.}
\label{dataCP12}
\end{figure}
To complete the peak picture we show the diagrams of the severe infected and the deaths for single day (Fig.11 e Fig.12), that are in agreement with the delay time $t_{d}$ of 4-5 days stimated numerically. Considering the uncertainty in performing swabs, the possible delays in processing swabs and the end of the incubation time (10-14 days) the picture of the peak in the temporal window 21-24 March seems to be reasonable.
\subsection{Comparison with Gompertz scenario and the estimation of lifes saved with LD}
\begin{figure}[hbt!]
\center\includegraphics[width=0.35\textwidth]{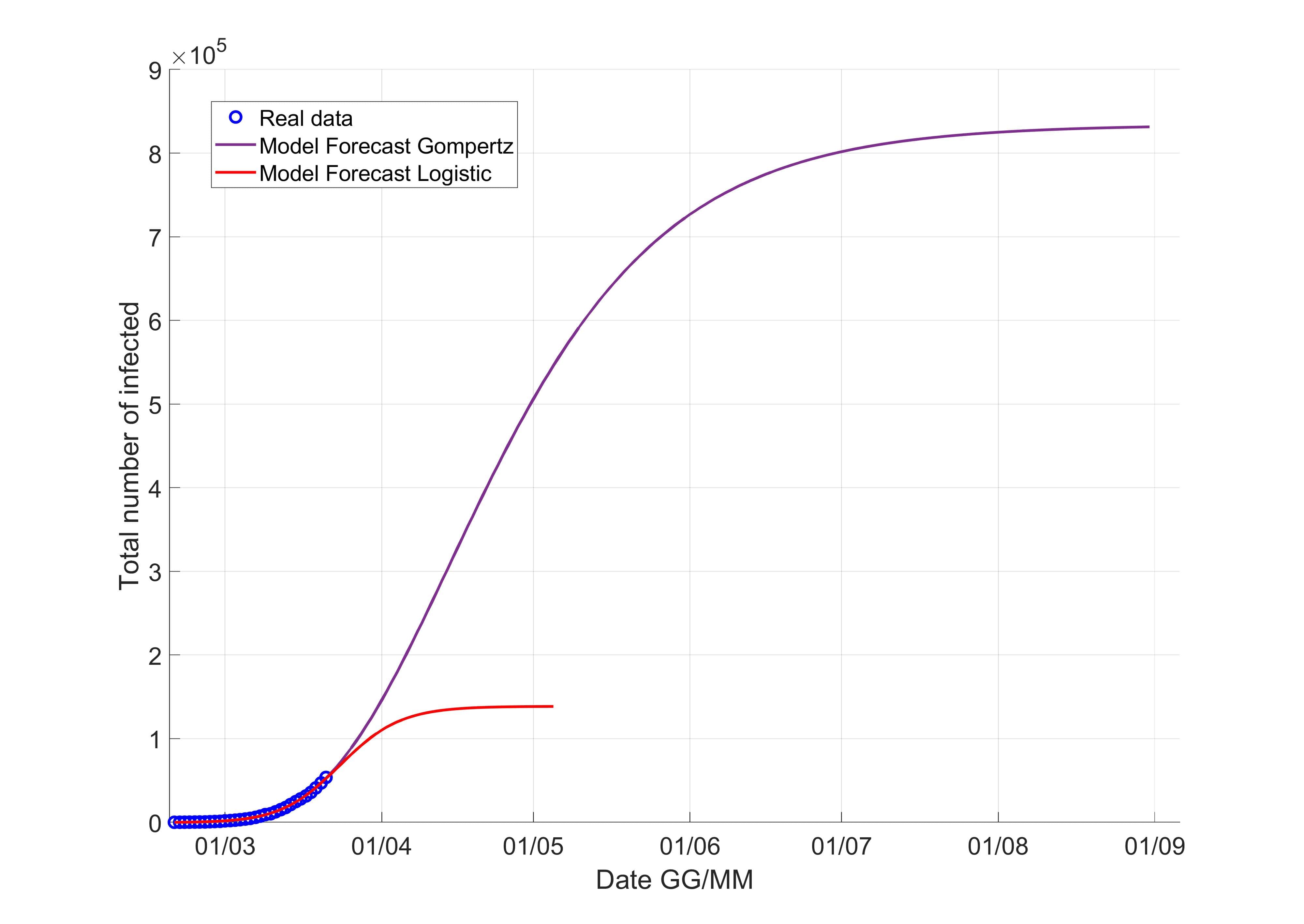}
\caption{Comparison between Gompertz and Logistic scenario}
\label{dataCP13}
\end{figure}
As we see in Fig. 13 the two scenarios are very different. We compare the generalized Logistic curve at 1 April (Fig. 8) with the Gompertz one calibrated with the data at 21 March ( i.e. with the data before the effect of the LD), imaging that the spread of the virus stops because the virus stops itself, i.e. the reproductive capacity of the virus decreases in time. Without LD we have this scenario:
\begin{itemize}
\item 831,000 severe infected,
\item 119,000 deaths,
\item 11,900,000 total infected,
\item the end of epidemic at early September.
\end{itemize}
This means that with LD we probably saved 100,000 human lifes!
\section{The situation for 5 regions}
Let's start with the analysis of the single regions. If we show the data of the Civil Protection in Fig.14 it's evident as any region has a velocity of propagation of the virus different from each other. In particular, Lombardia is very fast and Veneto is much slow at the beginning of the spread of the disease (we observed how in each region the number of days in which the number of infections rises by an order of magnitude ranges from 10 days in Lombardia to 16 in Veneto.).
\begin{figure}[hbt!]
\center\includegraphics[width=0.35\textwidth]{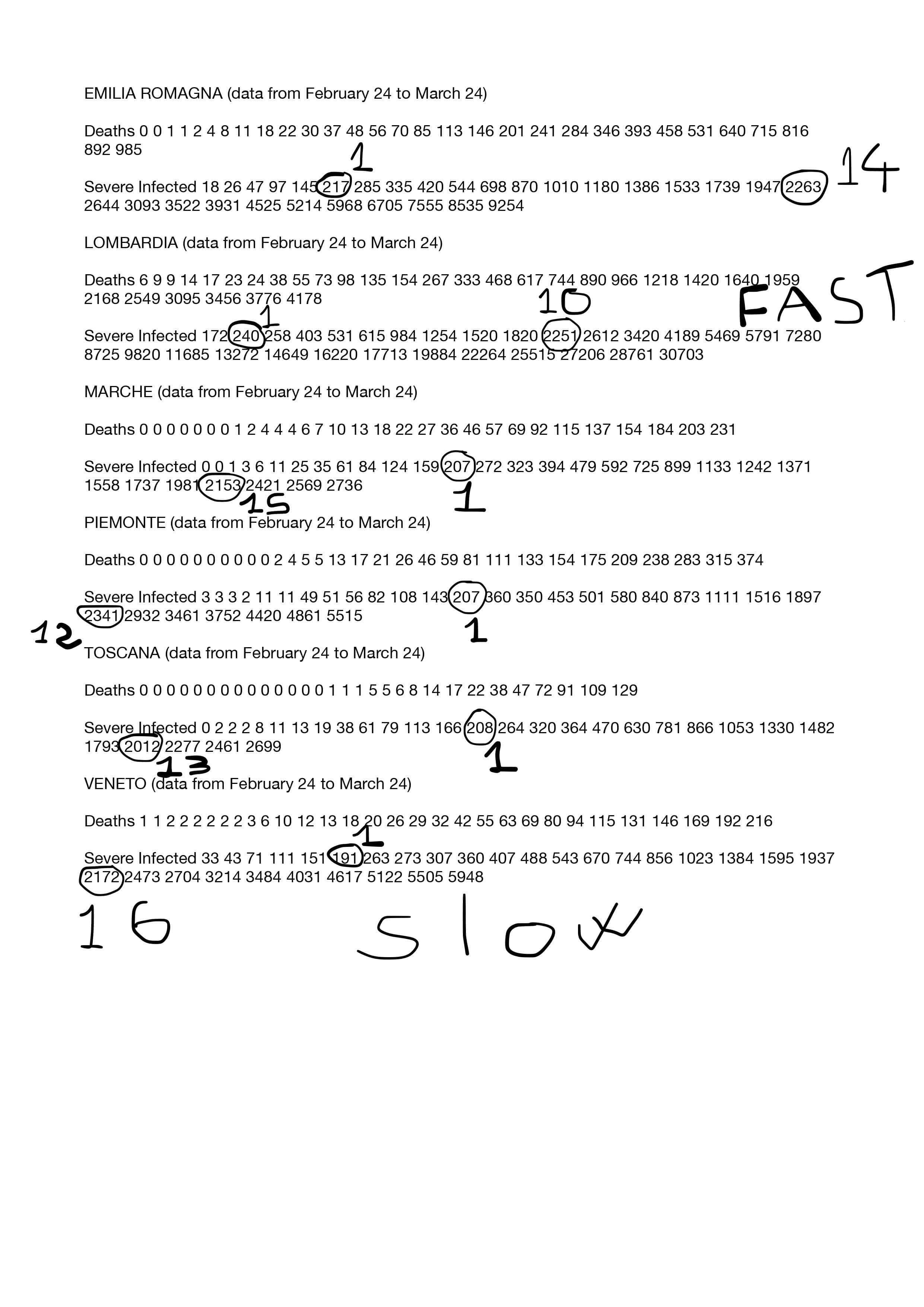}
\caption{The data of the Civil Protection for deaths and severe infected.}
\label{dataCP14}
\end{figure}
Looking at the data and from the previous simple observation, it is spontaneous to think that an analysis of the individual regions is more sensible than a global analysis of Italy. \\
This idea seems correct for two reasons: the virus arrives in the various regions at different times and each region has its different environmental and working characteristics, which mainly involves a different population density. After these considerations we therefore propose an analysis for the 6 regions mainly affected by the virus on 25 March.
\subsection{Lombardia Scenario}
Unfortunately, the forecast of Lombardia shows us that $59\%$ of future deaths are due to this region.\\
\begin{figure}[hbt!]
\center\includegraphics[width=0.35\textwidth]{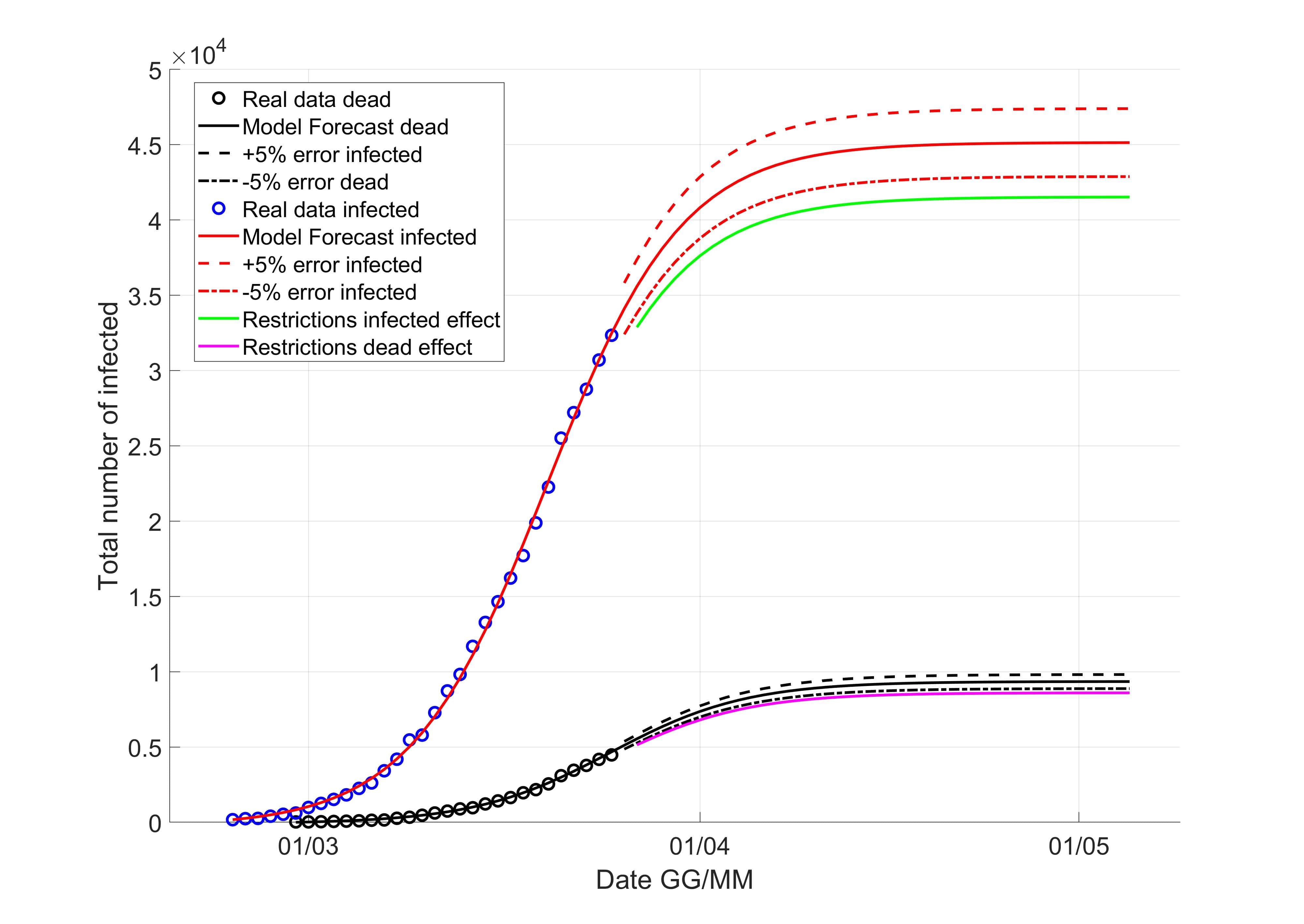}
\caption{The logistic curve of severe infected and deaths in Lombardia.}
\label{dataCP15}
\end{figure}
The parameter $A$ is the highest of all regions and this seems to confirm the interpretation that $A$ is the contribution of asymptomatics.\\
From the numerical simulations we obtain the following parameters
\begin{equation}
r_{0}=0.184, K=44900\quad\mbox{and}\quad A=49.2, 
\end{equation}
and
\begin{equation}
K_{1}=0.21\quad\mbox{and}\quad t_{d}=5. 
\end{equation}

\subsection{Emilia Romagna Scenario}
The region Emilia-Romagna will probably be the second most affected region in Italy. The geographical proximity to Lombardy and a large industrial activity could be the main causes.
\begin{figure}[hbt!]
\center\includegraphics[width=0.35\textwidth]{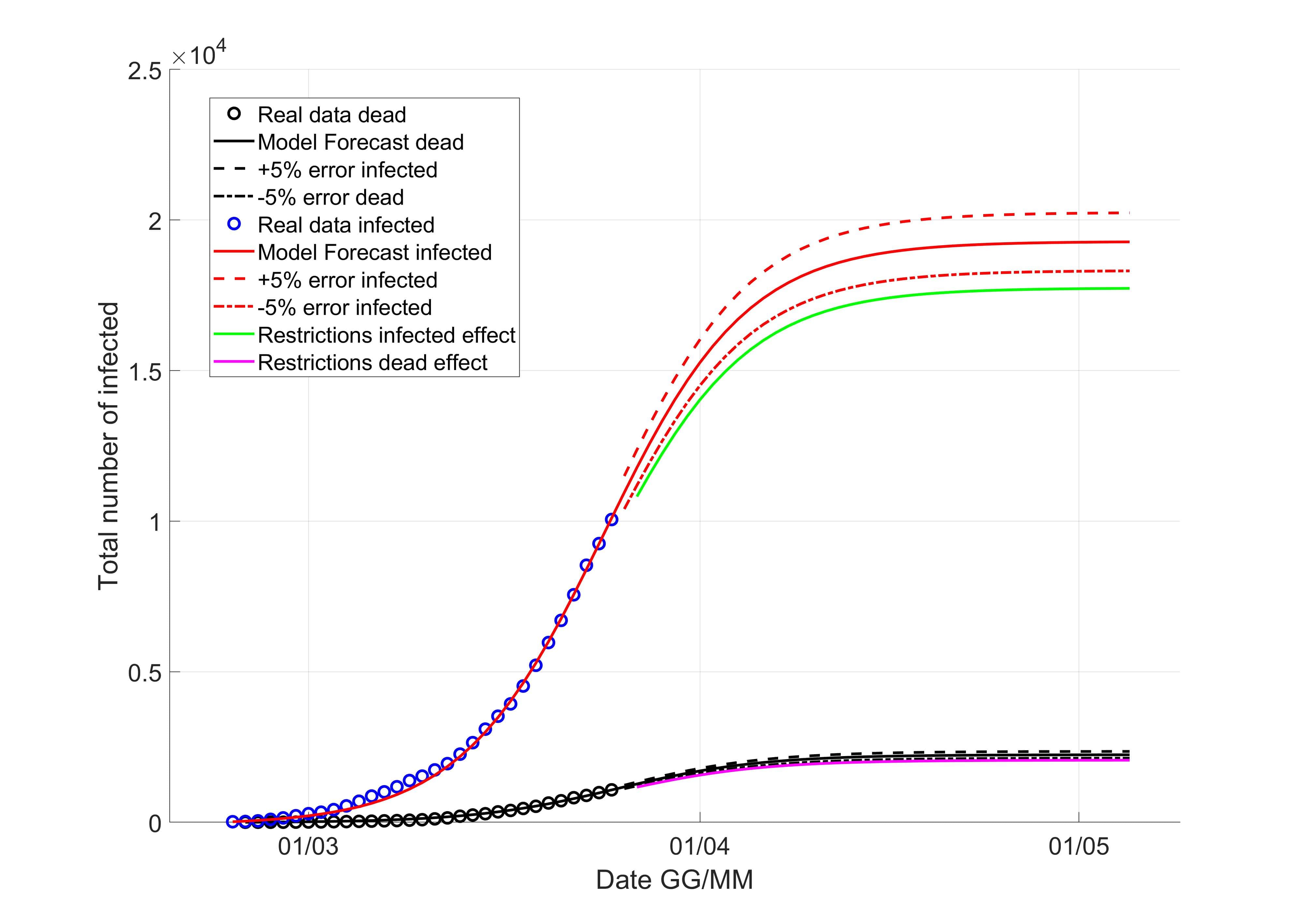}
\caption{The logistic curve of severe infected and deaths in Emilia Romagna.}
\label{dataCP16}
\end{figure}
From the numerical simulations we obtain the following parameters
\begin{equation}
r_{0}=0.175, K=19200 \quad\mbox{and}\quad A=15.5, 
\end{equation}
and
\begin{equation}
K_{1}=0.12\quad\mbox{and}\quad t_{d}=2.
\end{equation}

\subsection{Veneto Scenario}
For Veneto we make the opposite consideration compared to Lombardia: the coefficient A is the lowest among the regions that performed the largest number of swabs. \\
\begin{figure}[hbt!]
\center\includegraphics[width=0.35\textwidth]{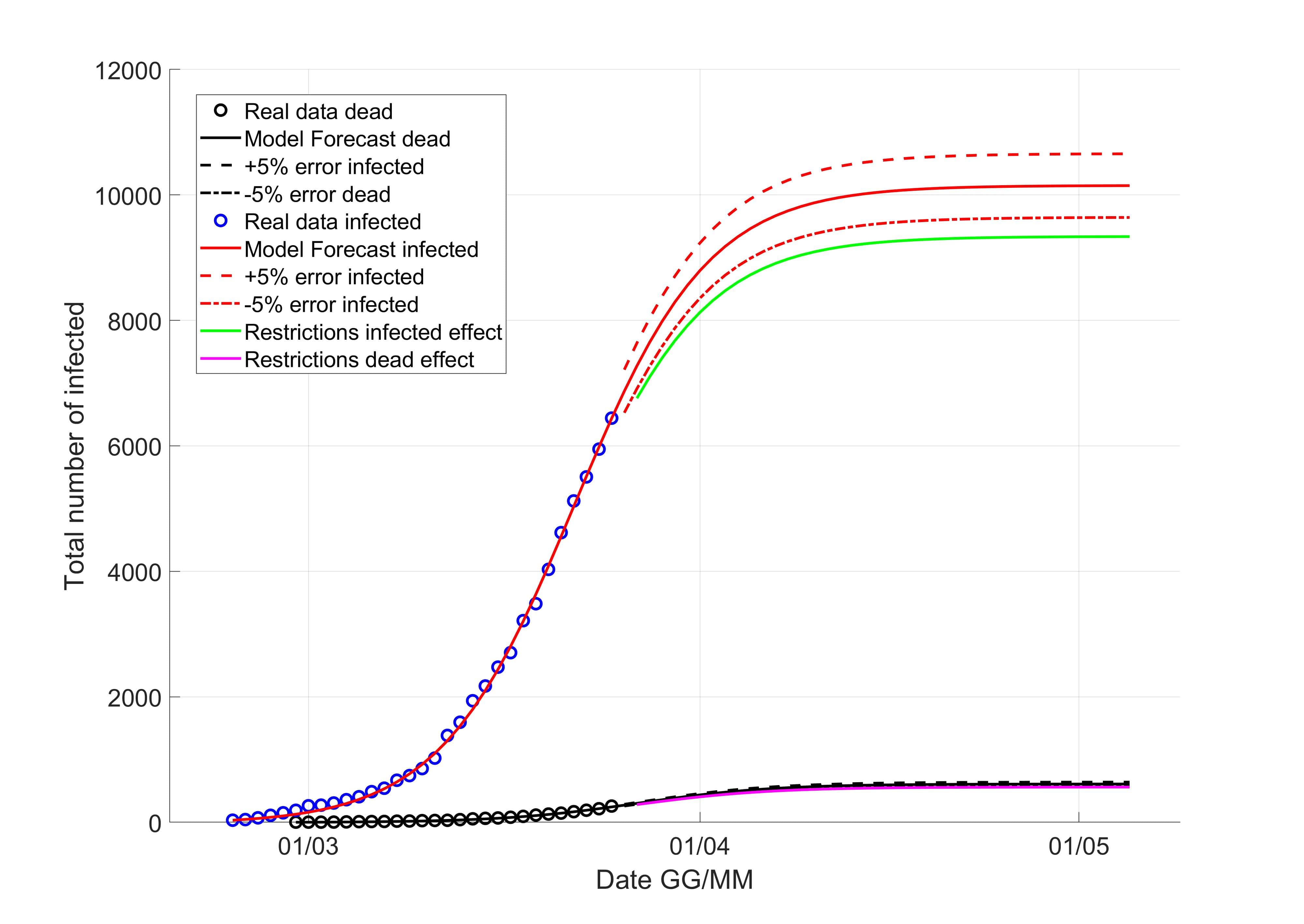}
\caption{The logistic curve of severe infected and deaths in Veneto.}
\label{dataCP17}
\end{figure}
This data seems to be in agreement with the fact that Veneto has carried out a large number of swabs also with asymptomatic so as to contain the spread of the virus by identifying potential vectors of the disease.\\ 
From the numerical simulations we obtain the following parameters
\begin{equation}
r_{0}=0.187, K=10100\quad\mbox{and}\quad A=5.4, 
\end{equation}
and
\begin{equation}
K_{1}=0.06\quad\mbox{and}\quad t_{d}=5.
\end{equation}
Yet another demonstration of our idea is contained in the $K_{1}$ value of the Veneto region, which is the lowest ever.
\subsection{Piemonte and Toscana Scenario}
\begin{figure}[hbt!]
\center\includegraphics[width=0.35\textwidth]{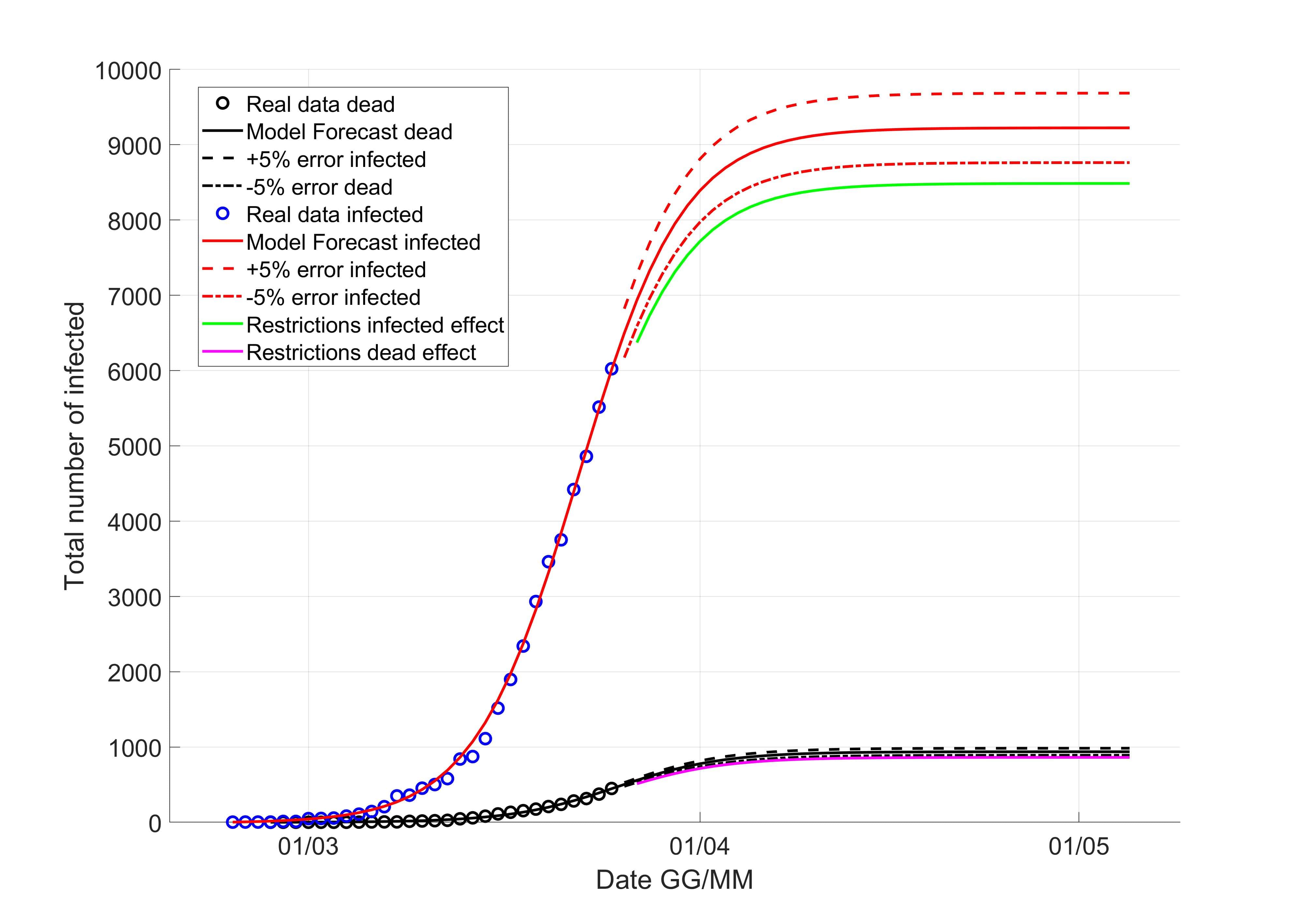}
\caption{The logistic curve of severe infected and deaths in Piemonte.}
\label{dataCP18}
\end{figure}
\begin{figure}[hbt!]
\center\includegraphics[width=0.35\textwidth]{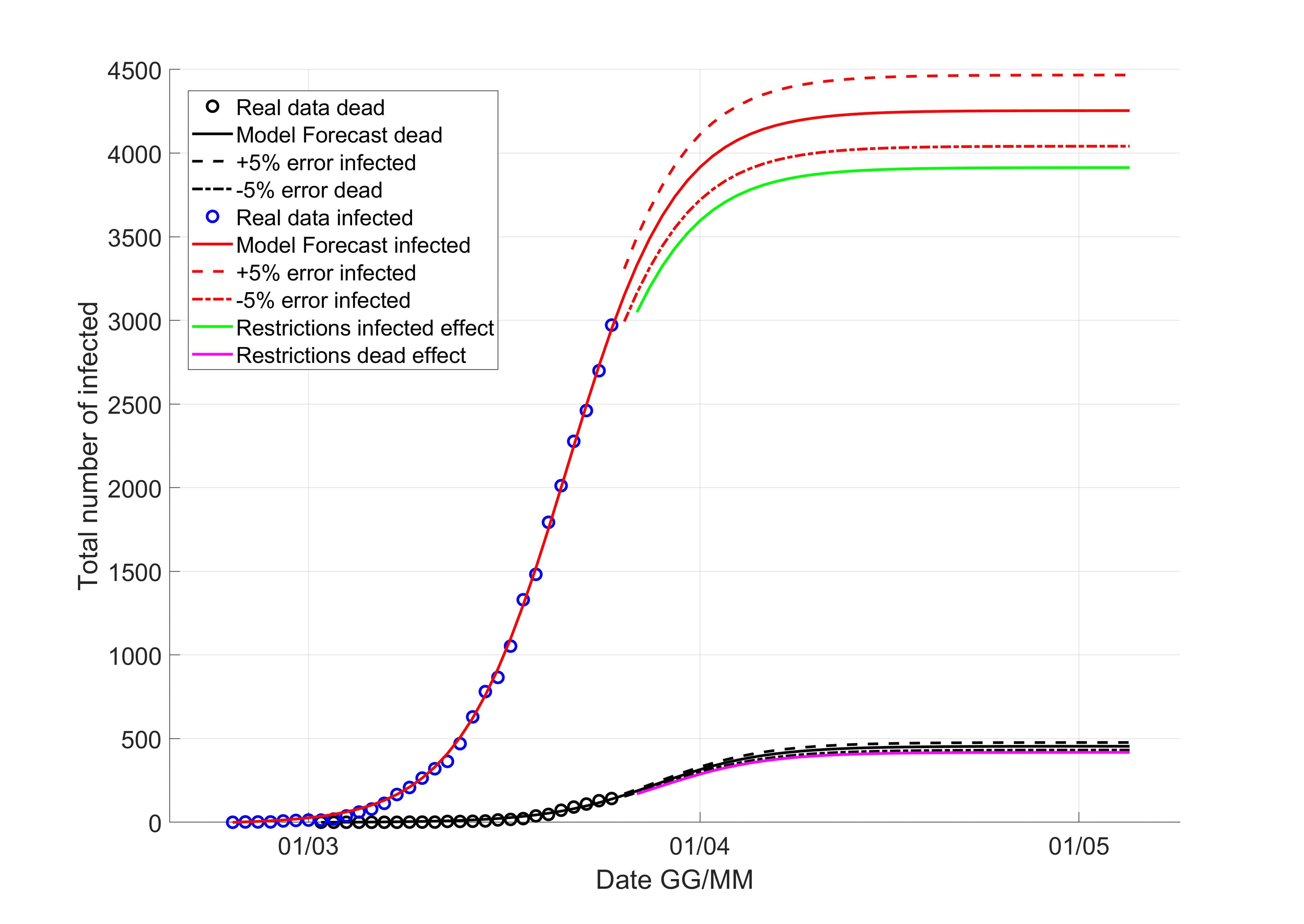}
\caption{The logistic curve of severe infected and deaths in Toscana.}
\label{dataCP19}
\end{figure}
From the numerical simulations we obtain the following parameters, respectively for Piemonte and Toscana
\begin{equation}
r_{0}=0.24, K=9200\quad\mbox{and}\quad A=2.57, 
\end{equation}
and
\begin{equation}
K_{1}=0.10\quad\mbox{and}\quad t_{d}=3.
\end{equation}
\begin{equation}
r_{0}=0.23, K=4250\quad\mbox{and}\quad A=2.05, 
\end{equation}
and
\begin{equation}
K_{1}=0.11\quad\mbox{and}\quad t_{d}=7.
\end{equation}
A brief comment about the numbers of these regions: the contagion in these regions certainly started later than the other 3. Then it could be that in the initial stages of the contagion the coefficient r is greater (as it was in our previous simulations for Italy) and then slowly going down, as you can see from Fig. 14 Piemonte and Toscana are quite fast. We justify the largest $r_{0}$ parameter for Piemonte and Toscana compared to Lombardia. We will do a sensitivity analysis of the parameters later.
\section{Upgrading the regions with the generalized Logistic equation at 1 April: a brief analysis of the single peaks region by region}
\begin{figure}[ht]
\center\includegraphics[width=0.35\textwidth]{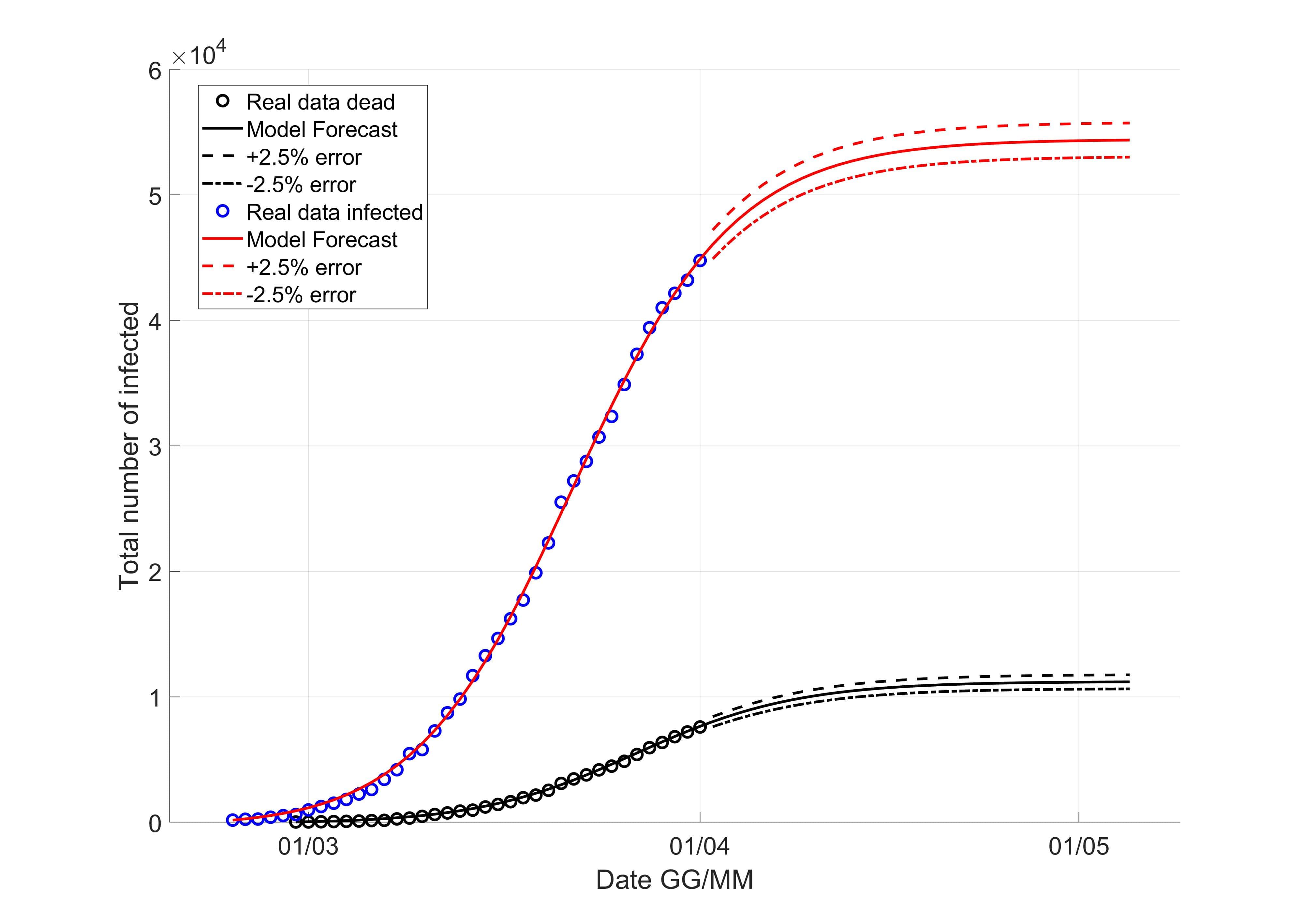}
\caption{The generalized logistic curve of severe infected and deaths in Lombardia at 1 April.}
\label{dataCP20}
\end{figure}
\begin{figure}[hbt!]
\center\includegraphics[width=0.35\textwidth]{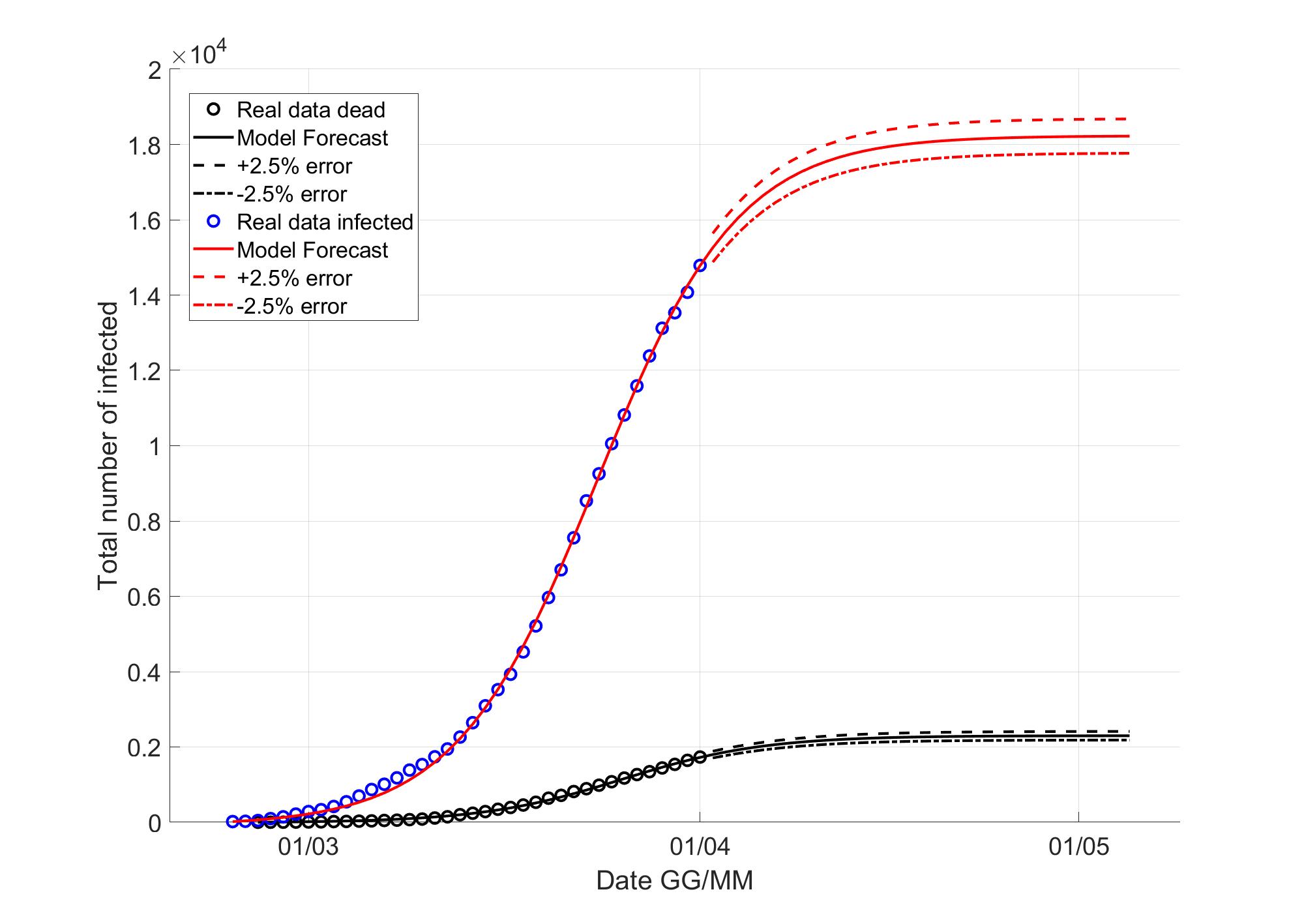}
\caption{The generalized logistic curve of severe infected and deaths in Emilia Romagna at 1 April.}
\label{dataCP21}
\end{figure}
\begin{figure}[hbt!]
\center\includegraphics[width=0.35\textwidth]{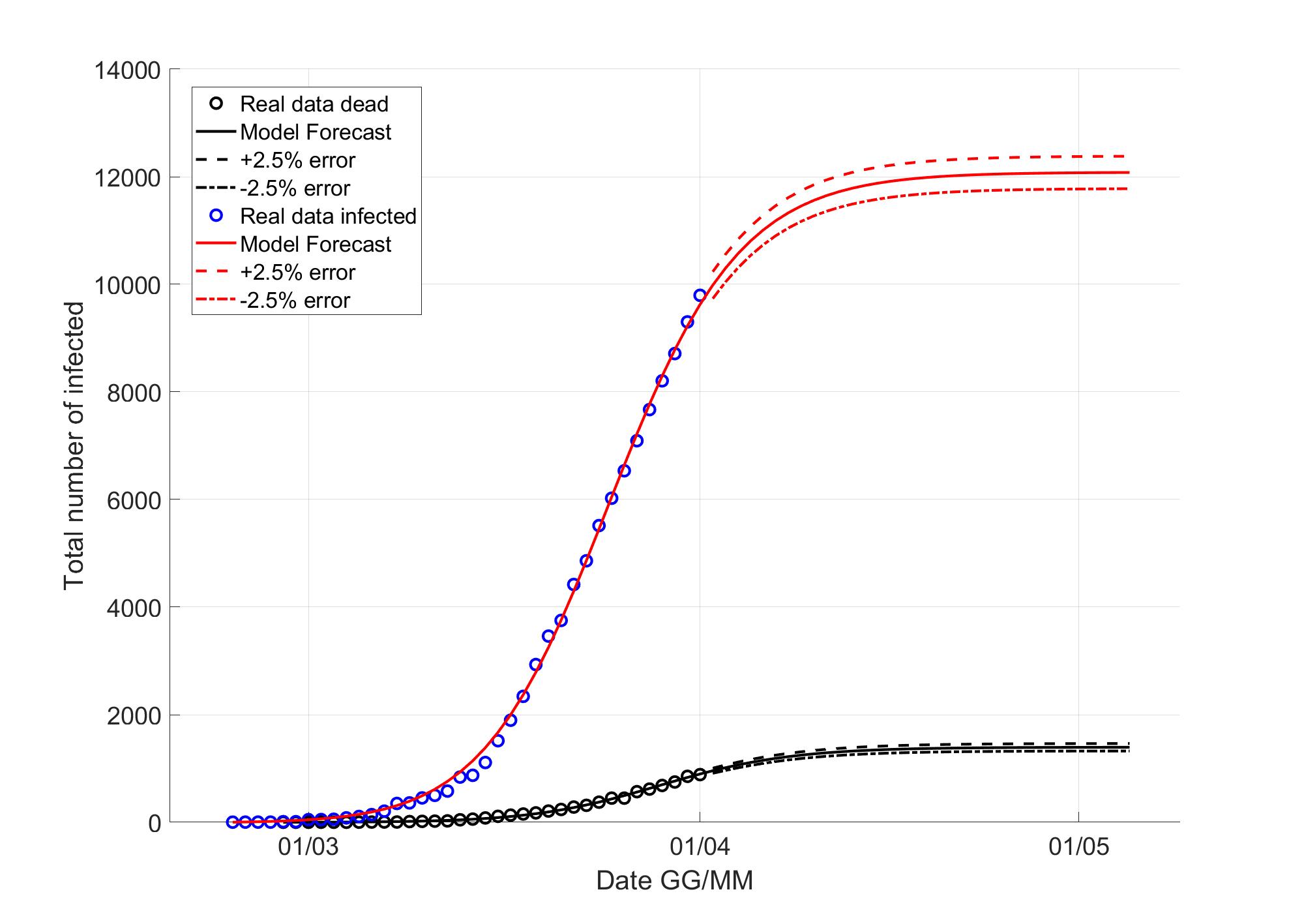}
\caption{The generalized logistic curve of severe infected and deaths in Piemonte at 1 April.}
\label{dataCP22}
\end{figure}
\begin{figure}[hbt!]
\center\includegraphics[width=0.35\textwidth]{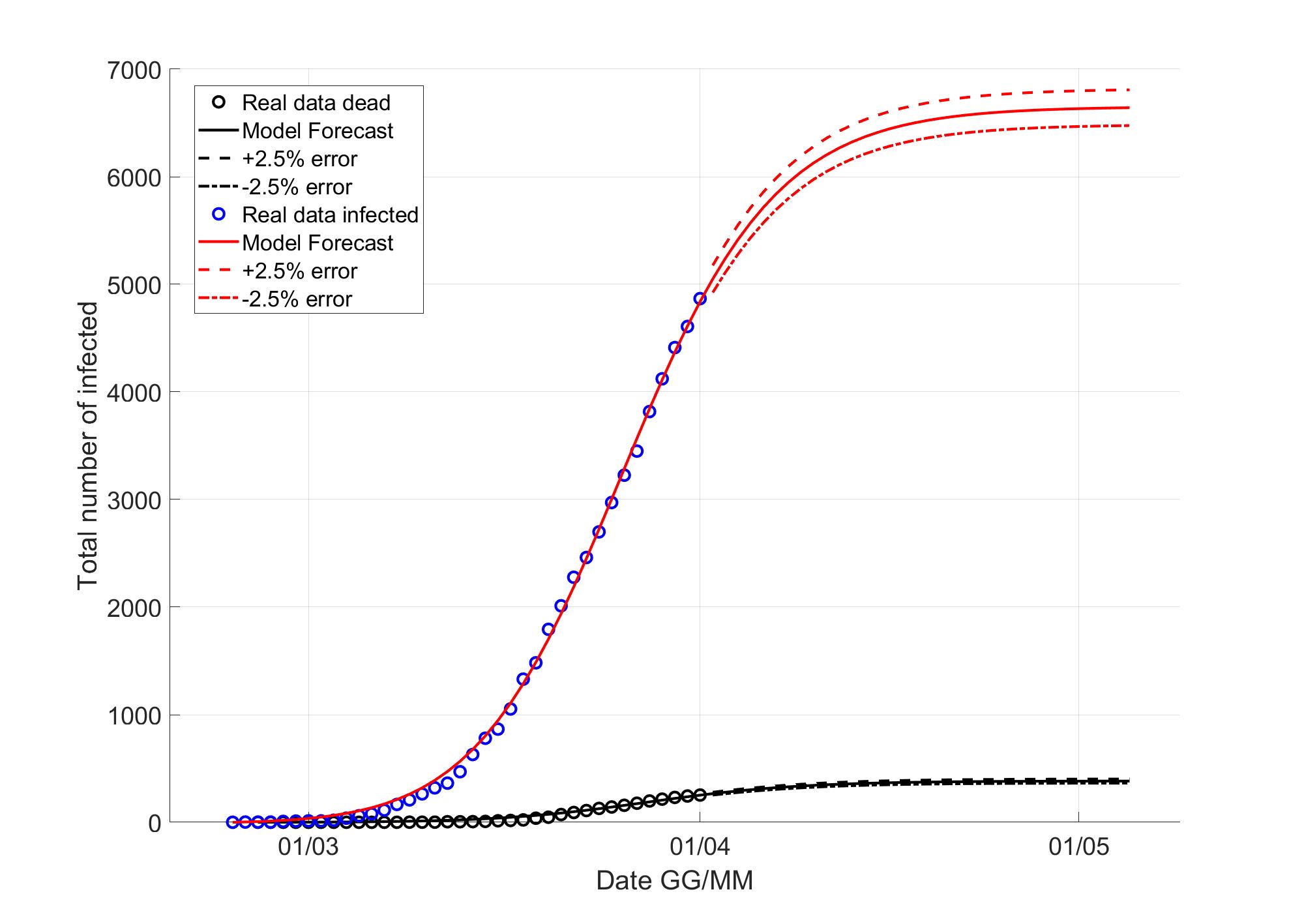}
\caption{The generalized logistic curve of severe infected and deaths in Toscana at 1 April.}
\label{dataCP23}
\end{figure}
\begin{figure}[hbt!]
\center\includegraphics[width=0.35\textwidth]{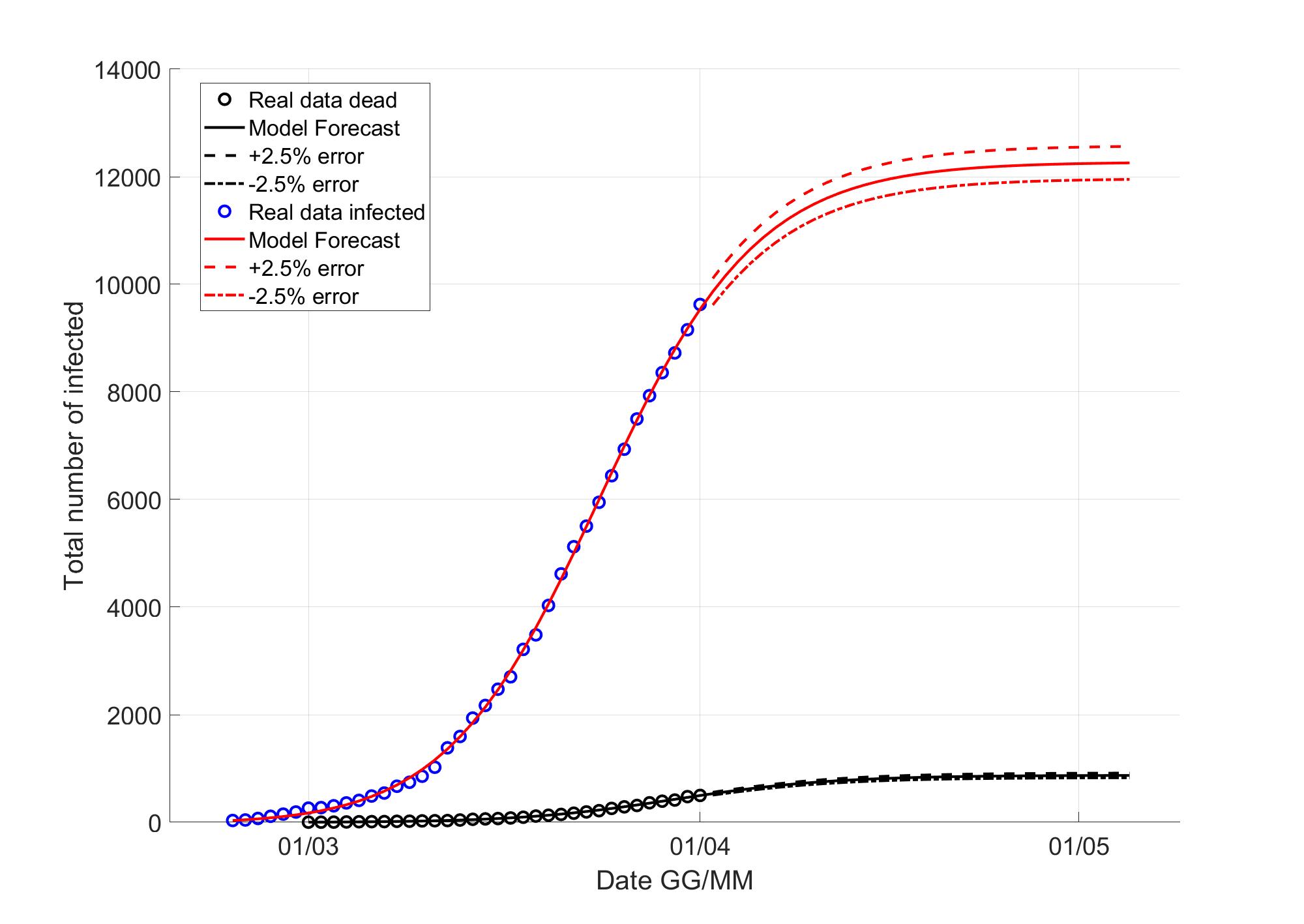}
\caption{The generalized logistic curve of severe infected and deaths in Veneto at 1 April.}
\label{dataCP24}
\end{figure}
\begin{figure}[hbt!]
\center\includegraphics[width=0.35\textwidth]{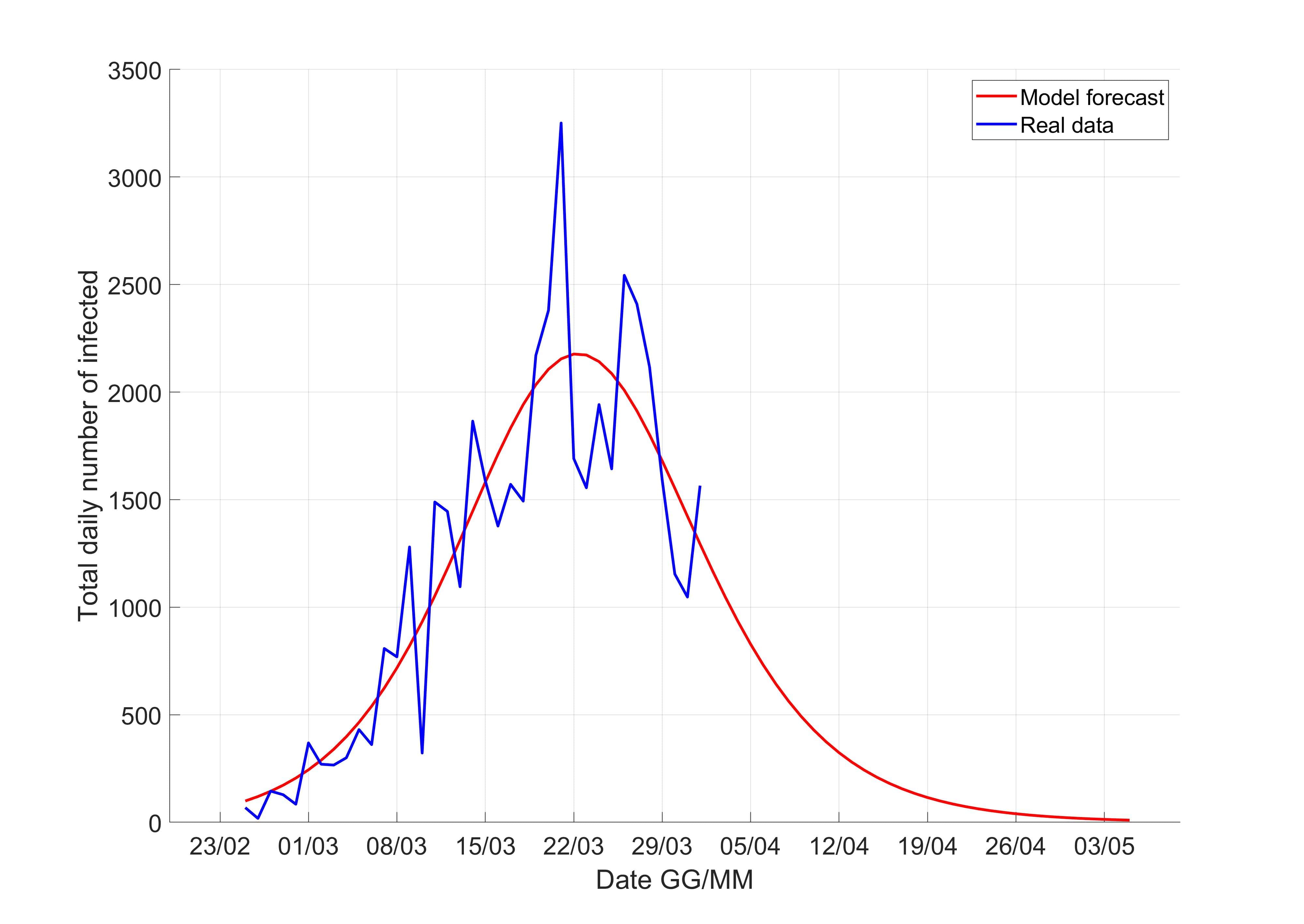}
\caption{The curve of daily severe infected in Lombardia at 1 April.}
\label{dataCP25}
\end{figure}
\begin{figure}[hbt!]
\center\includegraphics[width=0.35\textwidth]{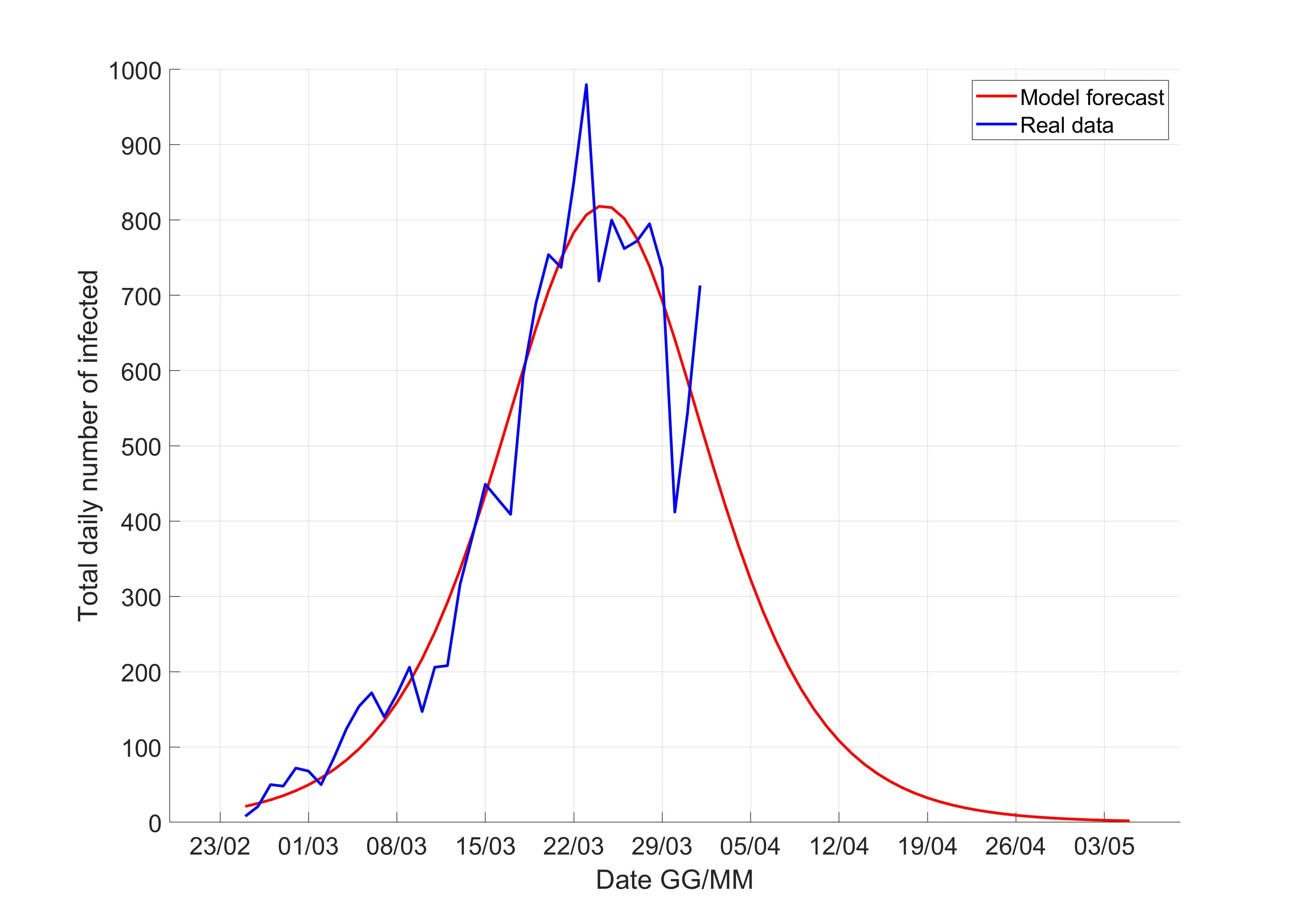}
\caption{The curve of daily severe infected in Emilia Romagna at 1 April.}
\label{dataCP26}
\end{figure}
\begin{figure}[hbt!]
\center\includegraphics[width=0.35\textwidth]{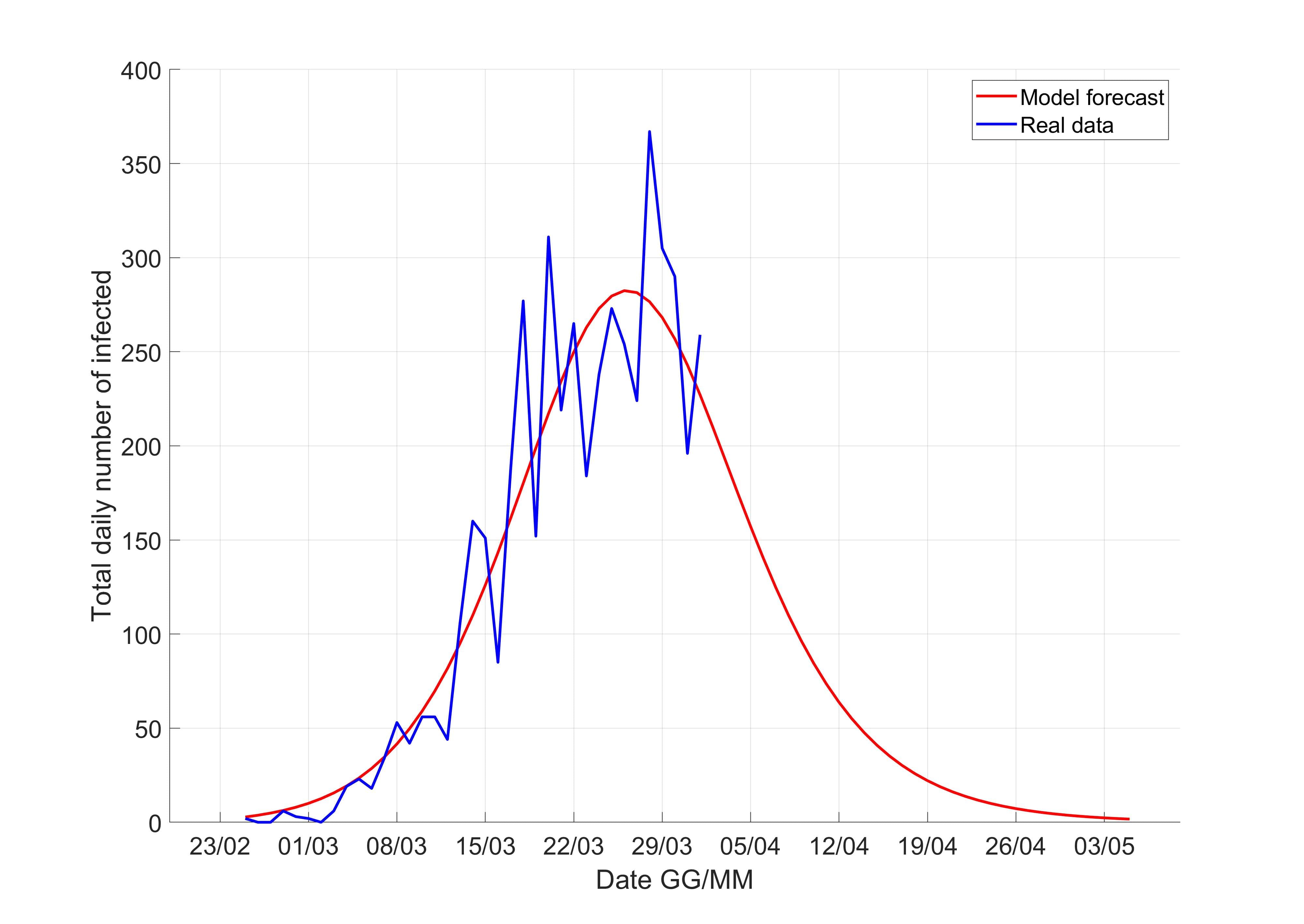}
\caption{The curve of daily severe infected in Toscana at 1 April.}
\label{dataCP27}
\end{figure}
\begin{figure}[hbt!]
\center\includegraphics[width=0.35\textwidth]{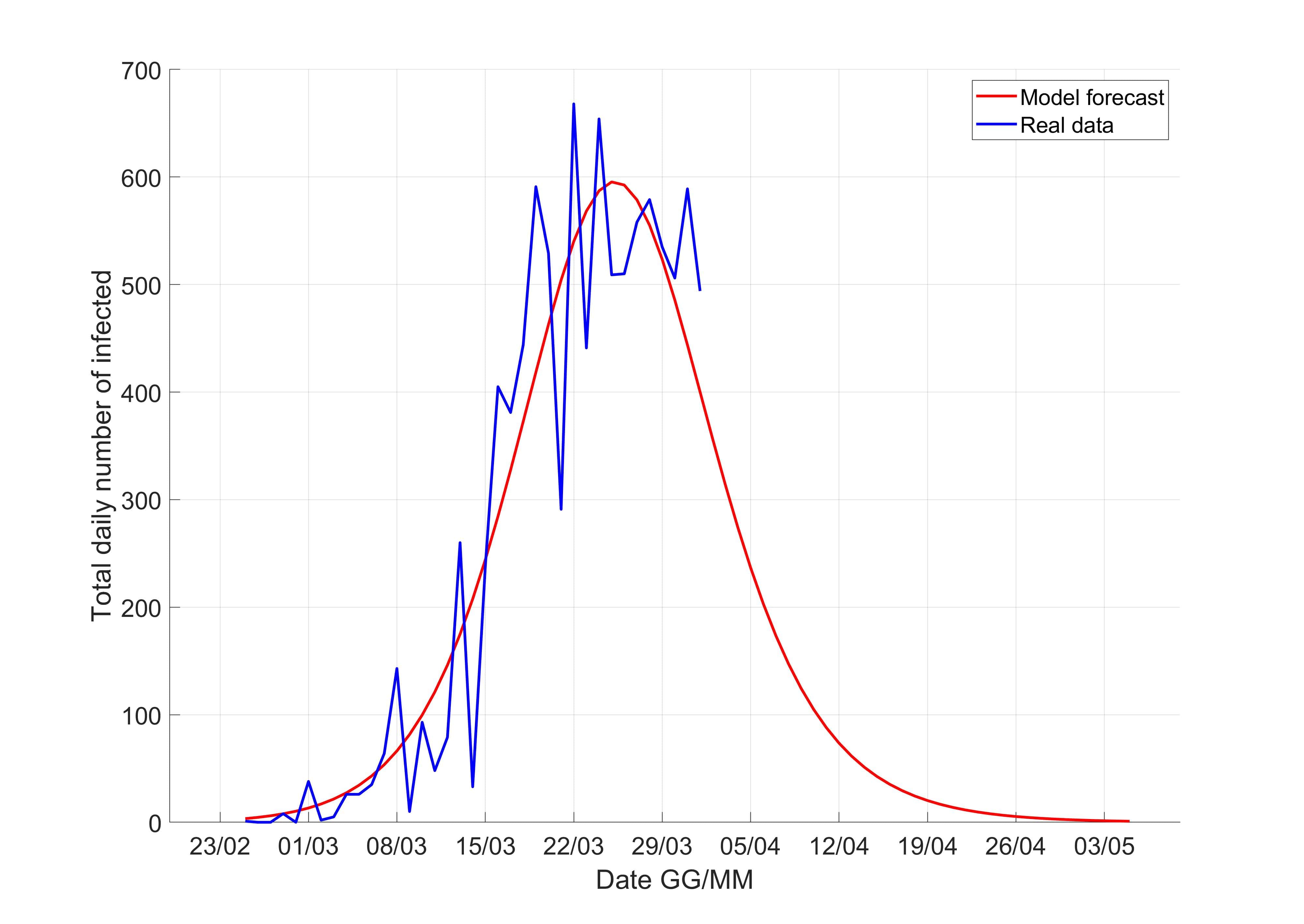}
\caption{The curve of daily severe infected in Piemonte at 1 April.}
\label{dataCP28}
\end{figure}
\begin{figure}[hbt!]
\center\includegraphics[width=0.35\textwidth]{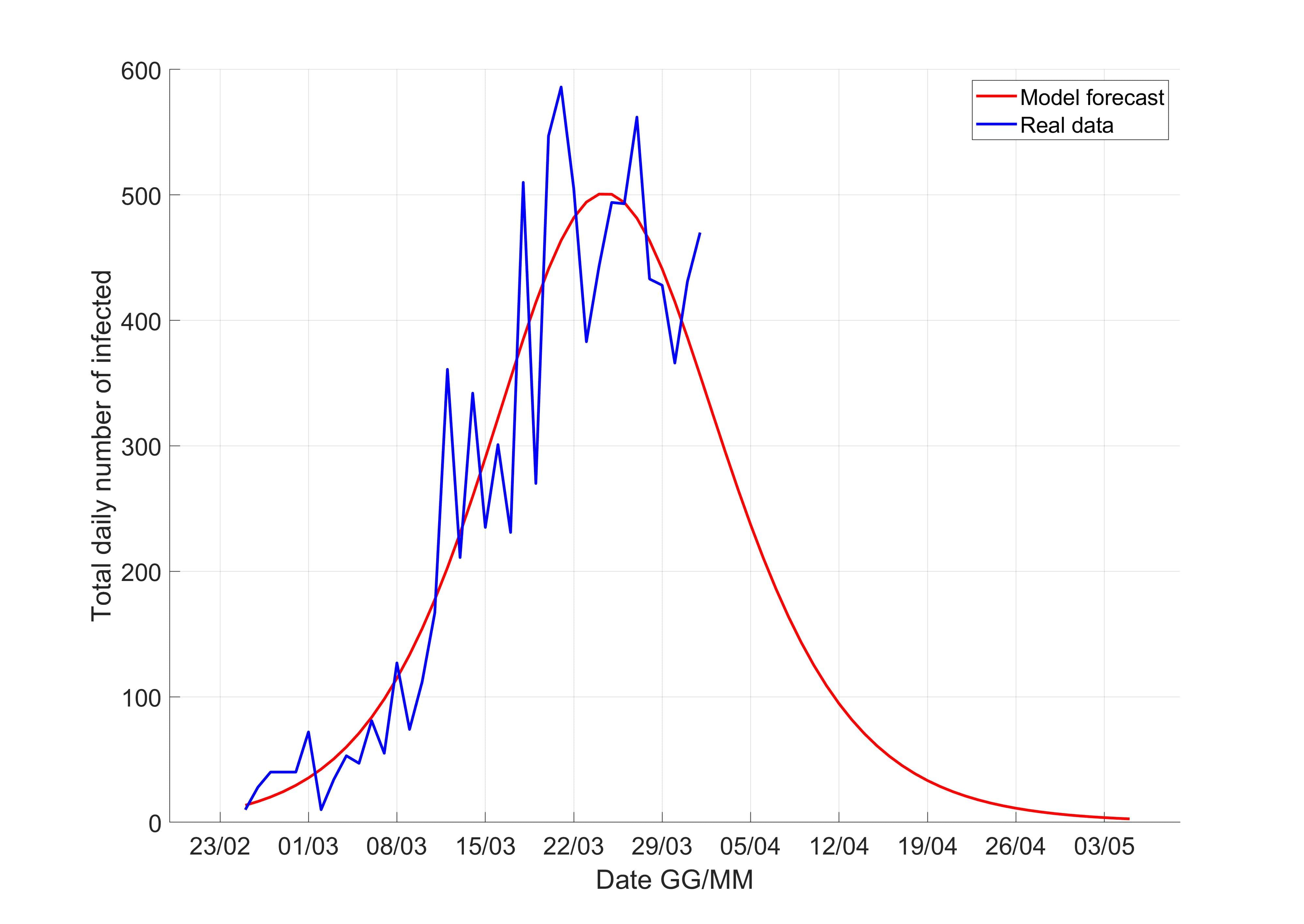}
\caption{The curve of daily severe infected in Veneto at 1 April.}
\label{dataCP29}
\end{figure}
As for the Italy case also for Lombardia and the other regions we have an increase of $20\%$ respect to the estimated values of 25 March as we see in Fig. 20-24, in fact we obtain
\begin{equation}
\nonumber
I(\mbox{end})=54370,
D(\mbox{end})=11190. 
\end{equation}
These numbers are very significative, it is the signal that Lombardia is the carrier region of the spread of the virus in Italy. \\ 
Finally as we see in Figs. 25-29 it is evident how the spread of the infection occurred at different times between the various regions, as the peaks are at different times. In particular, observing the tails of the curve, Toscana seems to be the region that will last come to zero infections, at 4-5 May.
\section{Conclusions}
In summary, we studied the evolution of Sars-Cov-2 in Italy with Civil Protection data in the time window from 24 February to 25 March with the Logistic model. We upgrade the curves of severe infected and deaths until 1 April with a generalized Logistic model, that works very well at the early of the contagion.\\
We provide around 138,5000 diagnosed patients and around 19,900 deaths on the whole Italian territory. This, considering a CFT of $1.2\%$, would lead to a total of about 1,658,000 infected at the end of coronavirus epidemic. We observe how the peak is more or less in correspondence of 21-24 March and the end towards the last days of April and early May as we see in Fig. 9. \\
We stress with the idea that a possible marker of the presence of the peak could be the quantity:
\begin{equation}
P_{\mbox{marker}}=\frac{I(t_{i})}{S(t_{i})}.
\end{equation}
This a quantity that has no fluctuations after passing its maximum value: for our opinion it is a great signal of stability.\\ Another important result is the estimate of the delay time $t_{d}$ between the peak of the severe infected and the peak of the deaths which seems to be truly in agreement with the data of the Civil Protection.
It is also interesting the comparison between the coefficient at 25 March and 1 April:
\begin{itemize}
\item $r_{0}=0.200$, $K=110950$ and $A=49$, at 25 March,
\item $r_{0}=0.173$, $K=138400$ and $A=79$, at 1 April.
\end{itemize} 
As it is obvious, the coeffecient $r_{0}$ is falling during the LD and the  the coeffecient $K$ is rising. We stress that it is very interesting that the contribution $A$ of asymptomatics is also increased. A possible interpretation is  that asymptomatics remain the most important sources of contagion in the single houses of italian people during the LD.
Finally we studied the evolution of the virus in the most affected regions. A delay is present in the appearance of the virus between the individual regions, so it seems more sensible to study them individually. We have noticed how Veneto's strategy in making swabs to as many people as possible has proved successful in containing the evolution of the virus.\\
In the upgrading of the single regions with the data at 1 April, we observe explicitly the peak of the daily severe infected and we show the effective time delay of the comparison of the Sars-Cov-2 in every regions. Finally Tuscany will be the last region with zero infections, at 4-5 May.\\
The last consideration about the importance of the LD: with a simulation with a Gompertz model, i.e. a model that predicts that the virus stops only because its infectivity propagation goes to zero, we estimate 119,000 deaths. With LD we have probably saved 100,000 human lifes!\\
Obviously this project is on-going being the possibility that the data of the next few days may change the scenario, hopefully positive, maybe in a further 6-7 days following the government's even more restrictive measures on 22 March.

\subsection{Acknowledgement}
We thank many colleagues for interesting discussions, in particular Andrea Marzolla, Domenico Seminara and Jacopo Viti, they were fundamental for correct observations and for providing us with sources from which to take inspiration. We also thank Pierluigi Blanc, S.O.C. Infectious Diseases 1 Santa Maria Annunziata Hospital, for stimulating discussions on technical subjects on which we had no knowledge.

\end{document}